\documentclass[sigplan,twocolumn]{acmart}
\renewcommand\footnotetextcopyrightpermission[1]{}
\usepackage{array} 
\usepackage{tikz}
\usepackage{amsmath}

\usepackage[]{hyperref}
\usepackage{amsfonts}
\usepackage{url}
\usepackage{filecontents}
\usepackage{xspace}
\usepackage{subfig}
\usepackage{multirow}
\usepackage{makecell}
\usepackage[ruled, linesnumbered, vlined]{algorithm2e}

\usepackage[T1]{fontenc}

\usepackage{enumitem}

\def\@mkbibcitation{\relax}

\begin{document}


\title[]{ \LARGE \sys{}: Generative Image Editing Made Efficient with Mask-aware Caching and
Scheduling}

\author{
    \rm{Xiaoxiao Jiang}$^{\dag *}$,
    \rm{Suyi Li}$^{\dag *}$,
    \rm{Lingyun Yang}$^{\dag}$,
    \rm{Tianyu Feng}$^{\dag}$,
    \rm{Zhipeng Di},
    \rm{Weiyi~Lu},
    \rm{Guoxuan~Zhu},
    \rm{Xiu~Lin},
    \rm{Kan~Liu},
    \rm{Yinghao~Yu},
    \rm{Tao Lan},
    \rm{Guodong Yang},
    \rm{Lin Qu},
    \rm{Liping Zhang},
    \rm{Wei Wang}$^{\dag}$
    \\
    $^{\dag}$Hong Kong University of Science and Technology
    \quad Alibaba Group
}

\newcommand{\sys}{\textsc{InstGenIE}}

\newcommand{\PHB}[1]{\noindent\textbf{#1}\hspace{.5em}} 
\newcommand{\PHM}[1]{\vspace{.4em} \noindent\textbf{#1}\hspace{.5em}} 

\newcommand{\secref}[1]{\S\ref{#1}}
\newcommand{\figref}[1]{Fig.~\ref{#1}}
\newcommand{\tabref}[1]{Table~\ref{#1}}
\newcommand{\thmref}[1]{Theorem~\ref{#1}}
\newcommand{\prgref}[1]{Program~\ref{#1}}
\newcommand{\algref}[1]{Algorithm~\ref{#1}}
\newcommand{\eqnref}[1]{Equation~\ref{#1}}
\newcommand{\clmref}[1]{Claim~\ref{#1}}
\newcommand{\lemref}[1]{Lemma~\ref{#1}}
\newcommand{\ptyref}[1]{Property~\ref{#1}}

\newcommand{\eg}{{e.g.\@\xspace}}
\newcommand{\ie}{{i.e.\@\xspace}}
\newcommand{\etc}{
        \@ifnextchar{.}
        \textit{etc}
        \textit{etc.\@\xspace}
}

\newcommand{\term}{\textsf}
\newcommand{\code}{\texttt}
\newcommand{\ths}{\textsuperscript{th}}
\newcommand{\circledtext}[1]{\raisebox{.5pt}{\textcircled{\raisebox{-.9pt} {#1}}}}

\newcommand{\todo}[1]{\noindent\textcolor{red}{[TODO: #1]}}
\newcommand{\todowriting}[1]{\noindent\textcolor{red}{[Writing: #1]}}
\newcommand{\todofigure}[1]{\noindent\textcolor{red}{[Figure: #1]}}
\newcommand{\todoexp}[1]{\noindent\textcolor{red}{[Experiment: #1]}}
\newcommand{\suyi}[1]{\noindent\textcolor{violet}{#1}}
\newcommand{\wei}[1]{\textcolor{red}{#1}}
\newcommand{\xiaoxiao}[1]{\textcolor{cyan}{#1}}

\setlength{\abovecaptionskip}{3pt plus 1pt minus 1pt}
\setlength{\belowcaptionskip}{3pt plus 1pt minus 1pt}
\setlength{\abovedisplayskip}{3pt}
\setlength{\belowdisplayskip}{3pt}

\newcommand{\TeaCache}{\textsc{TeaCache}\xspace}
\newcommand{\Ideal}{\textsc{Ideal}\xspace}
\newcommand{\Diffusers}{\textsc{Diffusers}\xspace}
\newcommand{\FISEdit}{\textsc{FISEdit}\xspace}
\begin{abstract}
Generative image editing using diffusion models has become a prevalent application in today’s AI cloud services.
In production environments, image editing typically involves a mask that specifies the regions of an image template to be edited. 
The use of masks provides direct control over the editing process and introduces sparsity in the model inference.
In this paper, we present \sys{}, a system that efficiently serves image editing requests.
The key insight behind \sys{} is that image editing only modifies the masked regions of image templates  
while preserving the original content in the unmasked areas.
Driven by this insight, \sys{}
judiciously skips redundant computations associated with the unmasked areas by reusing cached intermediate activations from previous inferences.
To mitigate the high cache loading overhead, \sys{} employs a \emph{bubble-free} pipeline scheme that overlaps computation with cache loading.
Additionally, to reduce queuing latency in online serving while improving the GPU utilization, \sys{} proposes a novel \emph{continuous batching} strategy for diffusion model serving, allowing newly arrived requests to join the running batch in just one step of denoising computation, without waiting for the entire batch to complete. As heterogeneous masks induce imbalanced loads,
\sys{} also develops a load balancing strategy that takes into account the loads of both computation and cache loading. 
Collectively, \sys{} outperforms state-of-the-art diffusion serving systems for image editing, achieving up to $3\times$ higher throughput and reducing average request latency by up to $14.7\times$ while ensuring image quality.
\end{abstract}


\maketitle

\def\thefootnote{*}\footnotetext{Equal contribution}

\section{Introduction}
Diffusion models are making significant strides in generative AI art, enabling the creation of high-quality, contextually accurate images. 
One of the daily-use image generation tasks is image editing, which modifies specific regions of an existing image template to achieve a desired outcome~\cite{couairon2023diffedit, huggingface_inpaint, meng2022sdedit}.
Image editing has a wide range of applications, from personal use to professional Photoshop~\cite{adobe_ps_fill,adobe_ps_editor}, and has fostered various use cases, such as virtual try-on~\cite{ootd,ootd_dataset}, face swapping~\cite{wang2024InstantID}, and image retouching~\cite{adetailor_blog}. 
Due to its wide applicability, image editing has matured into a service offered to users by Adode and Midjourney~\cite{adobe_image_edit, midjourney_editor}.
In a recent public diffusion model serving trace~\cite{katz}, 70\% of requests require image editing service to edit or retouch an image.
Similarly, in our production system, we collect a two-week trace that documents a large-scale image editing service using 20k GPU cards to generate 34 million images.

\begin{figure}[t]
  \centering
  \includegraphics[width=1.0\linewidth]{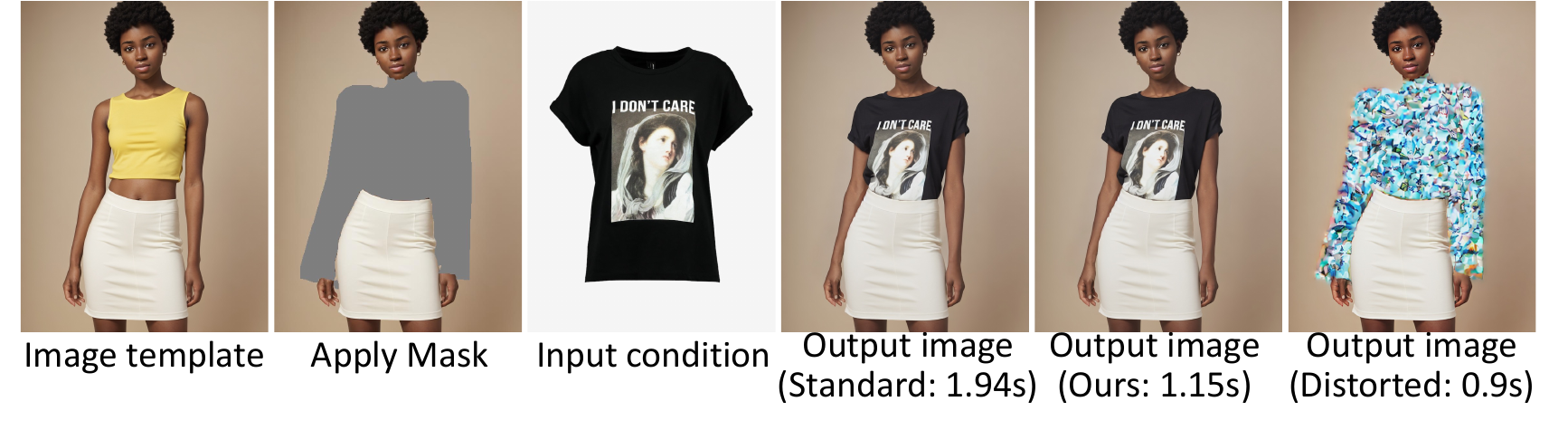}
  \caption{A virtual try-on example of image editing using a SDXL model on H800. \sys{} achieves a model inference speedup of $1.7\times$ and ensures image quality. 
  \textbf{The Rightmost image}: Naively disregarding unmasked regions in image editing will distort the output image.}
  \label{fig:edit_example}
  \vspace{-.2in}
\end{figure}

Typically, users employ a mask alongside other input conditions, such as textual prompts and images, to edit an image template. 
The use of a mask provides direct control, enabling users to precisely specify a region of arbitrary shape that they wish to modify while leaving the surrounding areas \emph{untouched}~\cite{ootd_dataset, couairon2023diffedit, kim2024stableviton, huggingface_inpaint,meng2022sdedit}. 
As illustrated in \figref{fig:edit_example}, the mask acts as a guide in the image editing process and is particularly favored by production services that demand accurate editing.
Notably, even if some image editing systems do not explicitly require users to provide masks, they will generate one based on other inputs, such as textual prompts, to facilitate the editing of a specific area in an image~\cite{fisedit, adetailor_blog}.
From our conversation with the production team, most image editing services require the mask guidance, which is either provided by the users or imposed by the platforms.

Despite the remarkable efficacy of image editing, 
serving their requests is challenging.
In existing systems, the computational complexity of 
editing an image is roughly equivalent to that of generating an entirely new image~\cite{diffusers,katz,li2024DistriFusion,liu2024timestep}.
This is because a diffusion model should model the relationships among all the images pixels in both the masked and unmasked regions to generate an image, using the attention mechanism~\cite{vaswani2017attention}:
naively disregarding the unmasked region to paint the masked region solely can distort the output image, as shown in \figref{fig:edit_example}.
Consequently, the requests will suffer from the 
high computational load of diffusion models, resulting in high inference latency and low serving throughput~\cite{li2024DistriFusion, katz, nirvana}.
For example, generating a 1024$\times$1024 image with the SDXL model~\cite{podell2024sdxl} requires 676T FLOPs~\cite{li2024DistriFusion}, saturating a high-end GPU like A100~\cite{katz}.
To address this challenge, existing approaches suggest leveraging multiple GPUs to accelerate the diffusion model inference~\cite{li2024DistriFusion,katz} or 
reusing intermediate activations in the inference process to skip computations during inference~\cite{nirvana, fisedit,liu2024timestep}.
However, employing multiple GPUs can negatively impact throughput, as existing work~\cite{li2024DistriFusion} achieves only a $2.8\times$ speedup using 8$\times$ more GPUs.
Worse, blindly skipping computation can degrade image quality ~\cite{katz}, which we will show in \S\ref{sec:eval_end2end_performance}.


In addition to the high computational load,
there has been limited attention given to the batching and routing of requests within diffusion model serving systems, resulting in a significant optimization gap. 
Existing systems~\cite{diffusers_server, katz, li2024DistriFusion} typically employ a \emph{static batching} policy~\cite{anyscale_continuous_batching}, where the running batch size remains fixed until its inference completes. 
As a result, requests that arrive while the model is executing inference cannot be processed until the current execution concludes, leading to prolonged queuing times that can increase tail serving latency by 35\% (\S\ref{sec:eval_continuous_batch}).
Moreover, blindly applying optimized strategies, such as continuous batching~\cite{vllm, yu2022Orca}, to image editing serving systems can yield suboptimal performance. 
Additionally, image editing requests vary in the size of utilized mask, as demonstrated by our characterization studies (\S\ref{sec:characterization}), and this heterogeneity
 should be considered by the request routing algorithm.


In this paper, we introduce \sys{}\footnote{Short for Instant Generative Image Editing}, an efficient serving system for generative image editing services. 
The key idea behind \sys{} is to avoid redundant computations in image editing by leveraging the guidance of the mask. 
As illustrated in \figref{fig:edit_example}, image editing modifies only the masked regions of the image while preserving the original content in the unmasked areas.
Following this insight, \sys{} accelerates the inference by caching and reusing the intermediate activations of the unmasked areas, thereby reducing the computational load.
Accelerating inference for requests further facilitates optimizations that enhance serving efficiency at the cluster scale, as the computation load of each request is reduced and multiple requests can be served in a batch.
In this context, \sys{} adapts the continuous batching strategy~\cite{yu2022Orca, vllm} to diffusion model serving and schedules requests judiciously to balance the load across multiple worker replicas. 
Following the design strategies,  \sys{} addresses three key technical challenges.

\textbf{First}, \sys{} accelerates image editing by reducing the computational workload associated with unmasked regions, focusing computation precisely on the masked regions.
\sys{} achieves this through \emph{mask-aware} image editing, which reuses the pre-computed activations from previous requests to provide global context and pixel interactions for the current request.
Specifically, for requests involving edits to an image template that has been previously processed, the intermediate activations associated with the common unmasked regions can be shared and reused, significantly reducing the computation load. 
In \S\ref{sec:characterization} and \S\ref{sec:key_insight}, we demonstrate the applicability of this approach to common image editing tasks by characterizing production workloads.
While reusing pre-computed activations can accelerate computation, caching them on the GPU is impractical due to their large size, often on the order of GiB.
Therefore, \sys{} stores the activations in host memory and employs a \emph{pipeline loading} scheme, which overlaps cache loading for unmasked tokens with computation for masked tokens. 
However, the latency of computation and loading can vary significantly due to the wide range of mask sizes used (\S\ref{sec:characterization}),
which can cause bubbles in a naive pipeline loading scheme, negatively impacting the inference latency. 
To tackle the challenge, we formulate the pipeline optimization as a dynamic programming problem to squeeze bubbles out and minimize inference latency (\S\ref{sec:cache_edit}).


\textbf{Second}, \sys{} improves serving efficiency using a novel continuous batching mechanism~\cite{yu2022Orca, vllm}, marking its \emph{first} application in diffusion model serving.
Compared to full image generation, the adoption of mask-aware acceleration significantly reduces the computational load per request and thus magnifies the performance gain of batching by $1.29\times$ with a batch size of 4, creating opportunities to leverage batching for higher throughput.
We observe that diffusion model computation features an iterative denoising process, where an image is generated through multiple denoising steps, e.g., 50~\cite{diffusers,katz,nirvana,li2024DistriFusion}.
Driven by this observation, we adapted the continuous batching design—originating from large language model (LLM) serving~\cite{yu2022Orca,vllm}—for image editing tasks, where 
completed requests immediately exit from the running batch after each denoising step and new requests can join the running batch in just one denoising step, without waiting for the entire batch inference to complete.
However, since diffusion model serving consists of both CPU-intensive image processing operations and GPU-intensive computations, naively applying continuous batching will interleave them~\cite{toppings, sarathi}, leading to suboptimal serving performance (\S\ref{sec:eval_continuous_batch}).
To tackle this challenge, \sys{} proposes a disaggregation method that separates CPU-intensive image processing from GPU-intensive denoising computation by distributing them to different processes, thereby preventing CPU operations from interfering with GPU computations.

\textbf{Third}, \sys{} incorporates a load balancing strategy to prevent hotspots within the cluster. 
Our workload characterization (\S\ref{sec:characterization}) reveals that the masks used in image editing requests differ in size vastly, which can introduce load imbalances among worker replicas if using \emph{mask-aware} acceleration for image editing. 
Simply dispatching requests uniformly across worker replicas—such as balancing based on the number of requests assigned to each server—is ineffective. 
To tackle the load imbalance, \sys{} proposes a \emph{mask-aware} load balancing strategy that takes mask size into account to assess a worker's load.
In specific, we develop regression models, fitted with the offline data, to estimate the latency of computing and cache loading.
By solving a dynamic programming problem, \sys{} can estimate each worker's load to make informed request routing decisions.



Putting it together, we prototype  \sys{} on top of HuggingFace Diffusers~\cite{diffusers} and evaluate it using real-world masks sampled from production traces.
Our evaluation incorporates three diffusion models that have different computational intensities, i.e., SD2.1~\cite{rombach2022high}, SDXL~\cite{podell2024sdxl}, and Flux~\cite{flux2024}.
We setup NVIDIA A10 and H800 GPUs to evaluate \sys{} and other baselines.
Evaluation results show that \sys{} outperforms the state-of-the-arts diffusion model serving systems, including Diffusers~\cite{diffusers}, FISEdit~\cite{fisedit}, and TeaCache~\cite{liu2024timestep}, 
achieving up to $3\times$ higher throughput and reducing average serving latency by up to $14.7\times$ while ensuring image quality.
\section{Background and Motivations}
\subsection{A Primer on Image Editing}
\label{sec:primer}

\begin{figure}[t]
  \centering
  \includegraphics[width=0.95\linewidth]{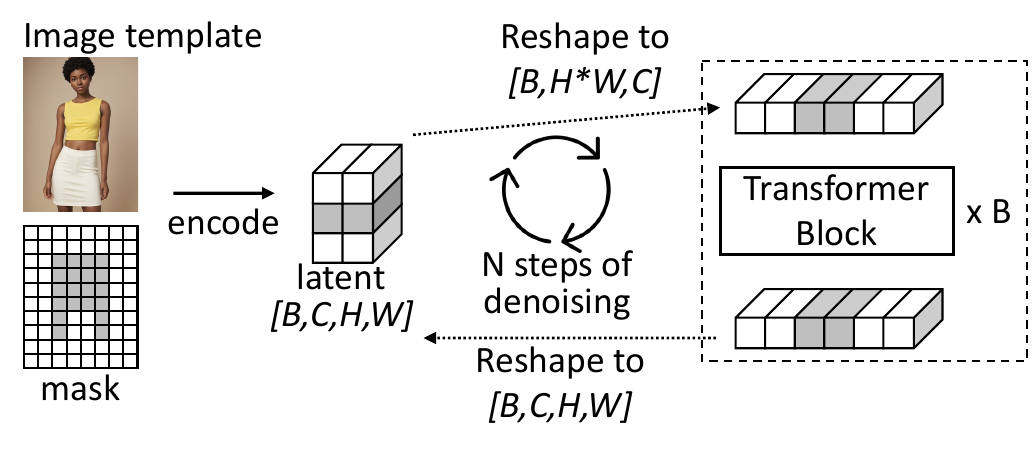}
  \caption{A simplified illustration of diffusion model inference. A darker cells/cuboid means it is masked.}
  \label{fig:pixel2token}
  \vspace{-.2in}
\end{figure}

Generative image editing with diffusion models is gaining popularity and has led to various applications such as virtual try-on~\cite{ootd}, face swapping~\cite{wang2024InstantID}, and image retouching~\cite{adetailer,adetailor_blog}. 
This process usually starts with an existing image---an \emph{image template}---in which users mask a specific area for editing. 
\figref{fig:edit_example} illustrates a virtual try-on example~\cite{ootd}, where users overlay clothing items onto model images to show how the garments would appear on them.

In \figref{fig:pixel2token}, we illustrate a simplified image editing process. 
Initially, the template image and the mask in pixel space are encoded into latent space. 
A diffusion model then uses the latent for $N$ steps of denoising computation, producing a final latent that is decoded into the output image. 
Key components of diffusion models~\cite{podell2024sdxl, sd3, flux2024} are multiple transformer blocks\footnote{For UNet-based models, e.g., SDXL~\cite{ootd}, transformer computations account for 82\%; diffusion transformer (DiT) models are a stack of transformers.}
, which primarily perform attention and feed-forward computations~\cite{vaswani2017attention}. 
During a step of denoising computation, a latent of shape $(B, C, H, W)$ is reshaped to $(B, H \times W, C)$
to pass through transformer blocks. 
This means the transformer receives input with a batch size of $B$,  token length of $H \times W$, and hidden dimension of $C$.
The attention mechanism in the transformer blocks captures contextual relationships among pixels to generate high-quality and contextually accurate images.
While \figref{fig:pixel2token} illustrates the process of a UNet-based model, i.e., SDXL~\cite{podell2024sdxl}, diffusion transformer (DiT) models~\cite{sd3,flux2024} follow a similar approach.

\subsection{Characterizing Image Editing Workloads}
\label{sec:characterization}

In this section, we characterize the generative image editing workloads using production traces.

\PHB{Prevalence.} 
Image editing services are crucial and pose real-world challenges~\cite{midjourney_editor, adobe_image_edit}, as evidenced by a recent public trace of image generation~\cite{katz}, where 70\% of requests involve image editing services for tasks like image restoration~\cite{adetailer}, virtual try-on~\cite{ootd, ootd_dataset} and image inpainting~\cite{huggingface_inpaint}.
Additionally, we collected a 14-day workload trace in January 2025, logging a large-scale face-swap service in our production system that utilized 20k GPU cards, generating more than 34M images.
We will release the trace for public access.



\PHB{The need for masks.} 
Masks play a crucial role in all image editing requests across both traces~\cite{katz,ootd_dataset, couairon2023diffedit, kim2024stableviton, huggingface_inpaint}. 
While image editing services typically allow users to provide their own masks, these masks can also be automatically generated when users do not specify them~\cite{adetailer, adetailor_blog}. 
For instance, in the image restoration task from the existing trace~\cite{katz}, which involves repainting hands or faces in newly generated images to correct distortions and enhance details, masks are automatically created using external tools like Adetailer~\cite{adetailer, adetailor_blog} to delineate the editing regions.


\begin{figure}[t]
  \centering
  \includegraphics[width=0.93\linewidth]{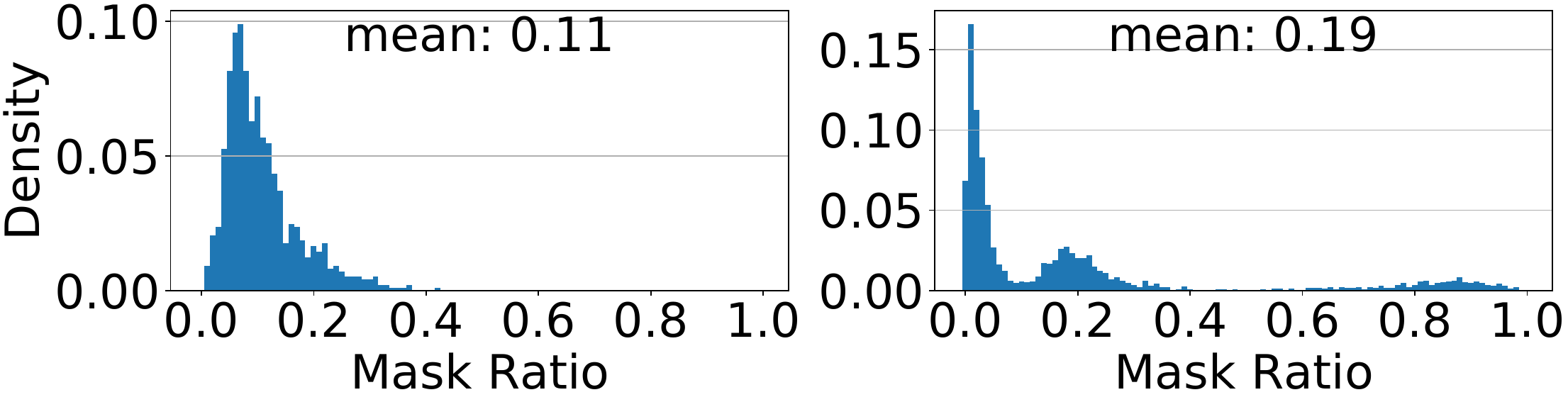}
  \caption{ 
  Mask ratio distributions of our traces (\textbf{Left}) and public trace~\cite{katz} (\textbf{Right}).
  }
  \label{fig:mask_ratio}
  \vspace{-.2in}
\end{figure}

\PHM{Masks differ in sizes and are generally small.} 
We analyze the \emph{mask ratios}---the proportion of masked area to total image area---in the traces~\cite{katz}.
As shown in \figref{fig:mask_ratio}, the average mask ratios are relatively small: 0.11 in our traces and 0.19 in the public trace. 
This indicates that editing regions are typically limited in size. 
We observe similar trends in another popular benchmark for virtual try-on~\cite{ootd_dataset}, with an average mask ratio of 0.35.
While the average is small, individual ratios exhibit a significant variation---meaning, the computation loads for requests can be vastly different, particularly if the editing inference process is \emph{mask-aware} as the computation can vary substantially with the specific masks.

\PHM{Reusability of the templates.}
Image editing tasks in the traces reveal that most requests involve modifying either existing image templates or newly generated images. 
In our trace, only 970 templates were utilized among the 34 million generated images, with each template being reused an average of 35,000 times. 
Similarly, the public trace~\cite{katz} shows that image restoration is immediately applied upon generating a new image. 
This observation suggests that for an image template to be editted, the intermediate activations of each pixel have likely been generated before and are available for reuse if possible.
In \S\ref{sec:key_insight}, we will analyze the activations associated with pixels generated by different image editing requests that target the same image template and discuss the feasibility of reusing these activations.


\begin{figure}[t]
  \centering
  \includegraphics[width=0.325\linewidth]{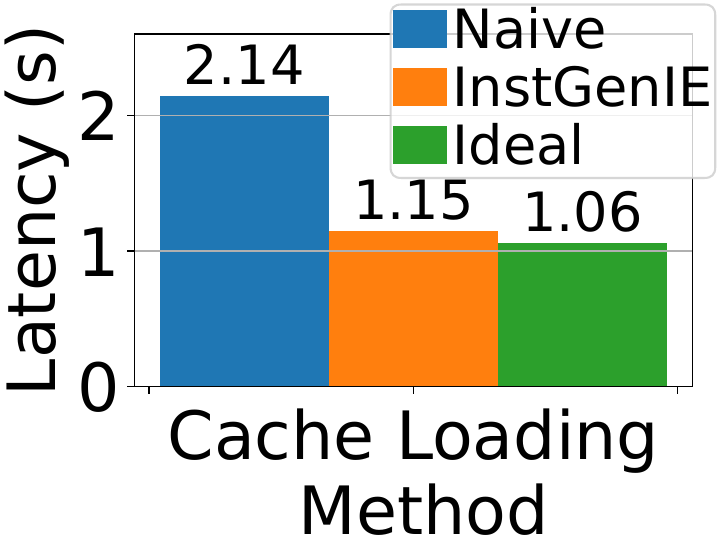}
    \includegraphics[width=0.325\linewidth]{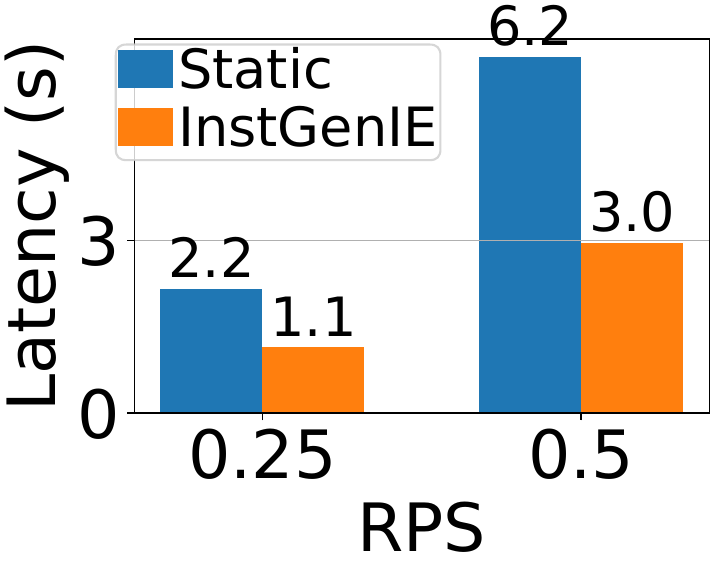}
      \includegraphics[width=0.325\linewidth]{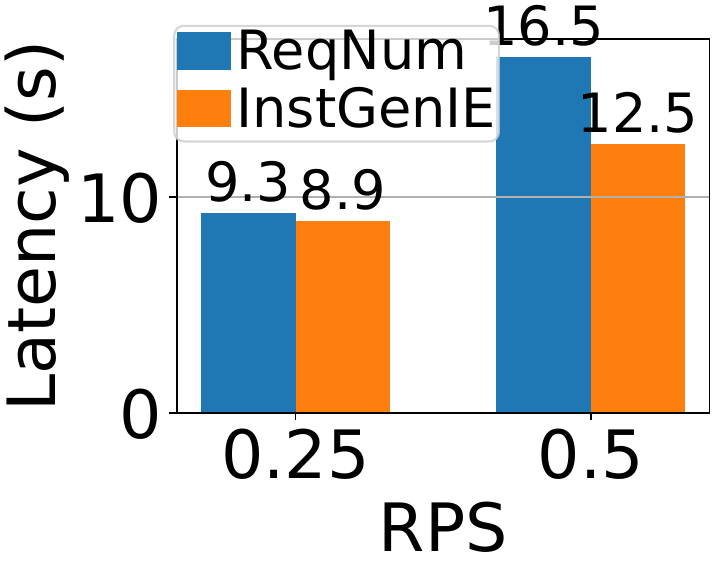}
  \caption{\textbf{Left}: Inference latency of a request using different cache loading methods. \textbf{Middle}: Queuing times undergone by requests with static batching~\cite{anyscale_continuous_batching} and \sys{}'s continuous batching under different requst traffic. \textbf{Right}: P95 tail latency of requests with naive load balance and \sys{}'s \emph{mask-aware} load balance. \textbf{RPS}: request per second.}
  \label{fig:motiv}
  \vspace{-.2in}
\end{figure}

\subsection{Opportunities and Challenges}
Our characterization studies highlight the widespread use of masks in image editing. 
Notably, for an image template, the masked regions are typically small, and the activations of the unmasked pixels are available from previous requests.
Driven by this observation, we propose the key insight that \emph{reusing activations associated with the pixels in the common unmasked regions can substantially reduce computational load}. 
Specifically, activations can be cached for reuse.
When a request edits an image template, the activations of the unmasked regions can be reused instead of recomputed, thereby accelerating inference. 
However, enabling mask awareness in the serving system poses significant challenges.


\PHM{C1: High cache loading overheads.} 
The primary goal of an image editing serving system is to achieve low serving latency for real-time user interaction~\cite{katz,nirvana}.
Given that the size of cached activations for an image template is on the order of GiB, storing them on GPU HBM is impractical. 
While host memory (DRAM) provides a feasible alternative, it incurs significant cache loading overhead.
As shown in \figref{fig:motiv}-Left,
naively performing sequential loading of activations from DRAM to HBM and executing inference can increase inference latency by 102\% for a SDXL model running on a H800 using PCIe Gen5~\cite{ootd}, compared with the ideal scenario where loading overhead is eliminated. 
\sys{} achieves performance comparable to the ideal case with its \emph{bubble-free} pipeline scheme, which effectively overlaps cache loading and computation (\S\ref{sec:cache_edit}).



\PHM{C2: Long queuing delay.}
Using cached activations instead of recomputing significantly reduces computational load: a single editing request can no longer saturate a GPU~\cite{katz,li2024DistriFusion}. 
This enables a unique opportunity to batch serving multiple requests
for enhanced throughput and GPU utilization. However,
naive static batching~\cite{anyscale_continuous_batching} in existing diffusion model serving systems~\cite{diffusers_server, katz, li2024DistriFusion} can result in 2$\times$ longer average queuing delays compared with \sys{}'s batching strategy, as shown in \figref{fig:motiv}-Middle, where we deploy a Flux model on H800.
This is because static batching does not allow new arrived requests to join the running batch on a worker until the inference of the running batch concludes.
Further, directly applying existing continuous batching strategies~\cite{vllm, yu2022Orca} yields suboptimal performance, increasing the tail latency by 40\%, which we will elaborate in \S\ref{sec:continuous_batching} and \S\ref{sec:eval_continuous_batch}.

\PHM{C3: Load imbalance.}
Our characterization studies show that masks vary in size, leading to a load imbalance problem among worker replicas if enabling \emph{mask-aware} image editing inference (\S\ref{sec:cache_edit}).
\figref{fig:motiv}-Right illustrates an experiment using Flux models on H800 GPUs, where a naive request-level load balancing strategy that uniformly assigns requests to workers is ineffective, increasing the P95 latency by 32\%. 
This highlights the need for a \emph{mask-aware} request scheduler that accounts for the impacts of mask size on image editing computations to route requests.




\subsection{Inefficiencies of Existing Works}
\label{sec:existing_works}

In this part, we briefly describe existing diffusion model serving systems and discuss why they cannot address the above challenges.
Existing diffusion model serving systems are mask-agnostic and produce edited images through \emph{full-image} generation~\cite{diffusers}.
Consequently, they suffer from long inference lantecy 
due to the high computational load of the involved diffusion models~\cite{katz,li2024DistriFusion,nirvana}.
While there have been tailored inference optimizations for diffusion models, such as leveraging multiple GPUs~\cite{li2024DistriFusion} for parallel model inference or reusing intermediate activations in the inference to skip computations~\cite{nirvana, liu2024timestep},
these optimizations target general image generation tasks and overlook the guidance of masks in image editing.
Naively applying these techniques can negatively impact serving throughput and image quality.
For example, DistriFusion~\cite{li2024DistriFusion} achieves a $2.8\times$ speedup using $8\times$ more GPUs.
Although skipping computations can reduce inference latency without requiring more GPUs, we show that this method can degrade image quality in image editing tasks (\S\ref{sec:eval_end2end_performance}).
Previous work also exploits sparse computation to accelerate diffusion model inference for image editing by only computing the activations for the masked region using speficially designed sparse kernels.
However, this method only applies to small-sized model, i.e., SD2.1~\cite{rombach2022high} and cannot serve requests with differernt mask ratios in a batch, leading to degraded serving performance (\S\ref{sec:eval_end2end_performance}).

In addition, existing systems primarily optimize diffusion model inference on a single server and often adopt a constant batch size of 1 due to the heavy computational load of the diffusion models~\cite{katz, diffusers, li2024DistriFusion, nirvana}. 
Due to its limited batching benefits~\cite{katz}, static batching strategy is employed~\cite{diffusers_server, anyscale_continuous_batching}, which can result in long queueing times and increase the tail latency of request serving by $35\%$ (\textbf{C2}) when a server handles multiple requests in a batch.
Besides, none of these systems addresses the issue of load imbalance (\textbf{C3}). 

\section{Mask-Aware Image Editing}




\subsection{Key Insight}
\label{sec:key_insight}
As discussed in \S\ref{sec:existing_works}, 
existing diffusion model serving systems perform full-image regeneration to edit an image, and thus suffer from high computational load.
To address the limitation,
\emph{an efficient serving system should exploit sparsity introduced by the masks to accelerate image generation without compromising image quality}.

Following this insight, we propose a \emph{mask-aware} serving system, which selectively reduces the computational load associated with the unmasked regions in an image template to accelerate the image editing process.
As discussed in \S\ref{sec:primer}, pixels in an image are mapped as tokens for transformer block computation.
Leveraging the mask, we can categorize the tokens as \emph{masked tokens} and \emph{unmasked tokens}, allowing us to precisely differentiate their computations in the transformer blocks.
For an image template, its pixels corresponding to the unmasked tokens are supposed to be untouched.
Intuitively,
intermediate activations generated during inference computation that are associated with these unmasked tokens can be cached and reused in subsequent requests that edit the same template, thereby eliminating the need for re-computation.

\begin{figure}[t]
  \centering
  \includegraphics[width=0.99\linewidth]{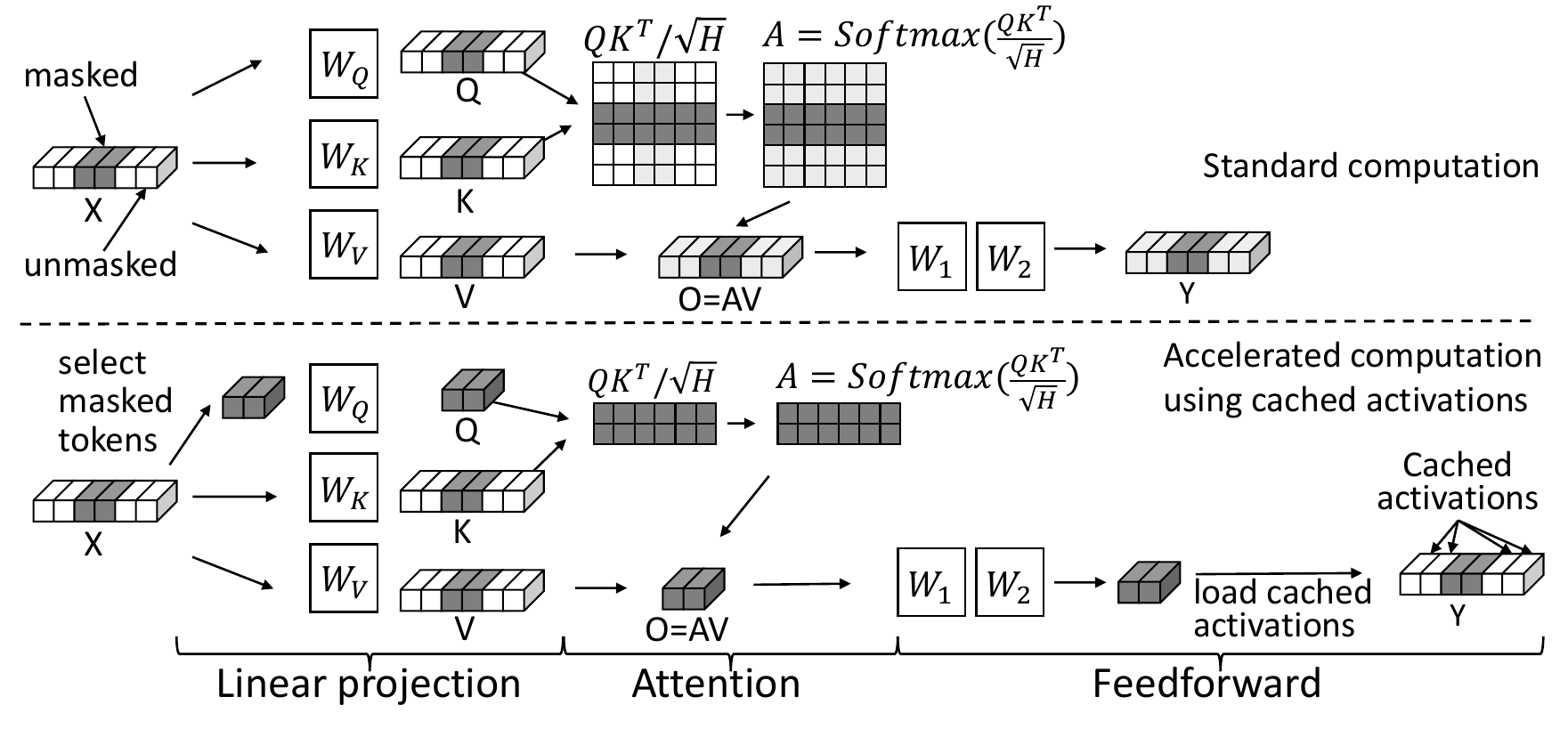}
  \caption{Main computations in a transformer block. 
  A darker cell/cuboid means it contains more information about the masked tokens.
  We omit LayerNorm~\cite{layernorm}, GeLU, and dropout for simplicity, which will not affect the results.}
  \label{fig:transformer}
  \vspace{-.2in}
\end{figure}

\PHM{How does it work in \sys{}?}
In \figref{fig:transformer}, we show the main computation operators in a transformer block and compare the standard computation flow with that of \sys{}.

Tokens in transformer blocks are discrete, and their computations are generally independent, except during the attention computation~\cite{vaswani2017attention}, where the computations of $\mathbf{QK}^T$ and \texttt{softmax($\cdot$)} introduce inter-token dependencies, i.e., the computation results rely on the values of multiple tokens.
Other computations, including linear projection, feedforward, LayerNorm~\cite{layernorm}, GeLU activation, and dropout, are token-wise, meaning that the computation of each token occurs independently of the others.
Consequently, \emph{for these token-wise operations, we can precisely differentiate the computations of the masked tokens and the unmasked tokens.}



\PHM{Mask-aware attention.}
\figref{fig:transformer}-Top illustrates the standard computation process of a transformer block.
We start with an input $\mathbf{X} \in \mathbb{R}^{B \times L \times H}$, where $B$ is the batch size, $L$ is the token length, and $H$ is the hidden dimension.
Some tokens within $\mathbf{X}$ are masked.
First, a linear projection maps $\mathbf{X}$ into $\mathbf{Q}$, $\mathbf{K}$, $\mathbf{V}$, using the weight matrices $\mathbf{W_Q}$, $\mathbf{W_K}$, $\mathbf{W_V}$, respectively.  
As the linear projection computation is token-wise, the computations for masked and unmasked tokens are independent.
Subsequently, the scaled matrix multiplication $\mathbf{QK^T}/\sqrt{H}$ combines $\mathbf{Q}$ and $\mathbf{K}$, during which the values of masked and unmasked tokens are multiplied according to the rule of matrix multiplication.
This results in some entries in the resulting matrix being derived from both masked and unmasked tokens (indicated by lighter gray cells).
Then, the \texttt{softmax} function is applied row-wise to $\mathbf{QK^T}/\sqrt{H}$, producing $\mathbf{A}$,
where all elements in $\mathbf{A}$ will be derived based on the value of masked tokens.
While the subsequent computations $\mathbf{O} = \mathbf{AV}$ and feedforward are token-wise, 
the activations in the output $\mathbf{Y}$ corresponding to the unmasked tokens are indirectly affected by the values of the masked tokens.


\begin{figure}[t]
  \centering
  \includegraphics[width=0.57\linewidth]{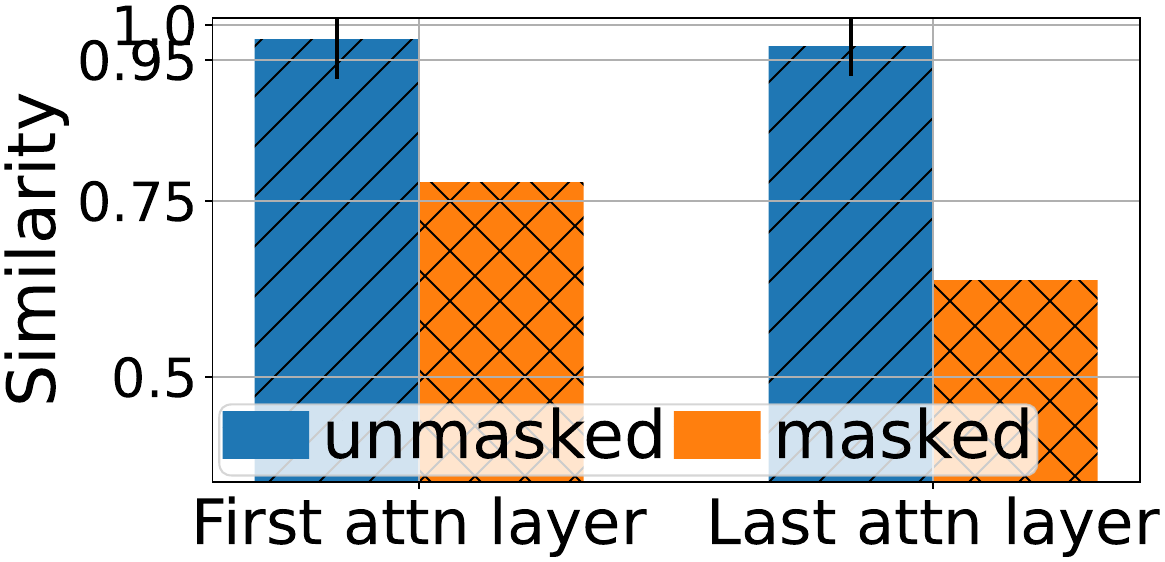}
  \includegraphics[width=0.42\linewidth]{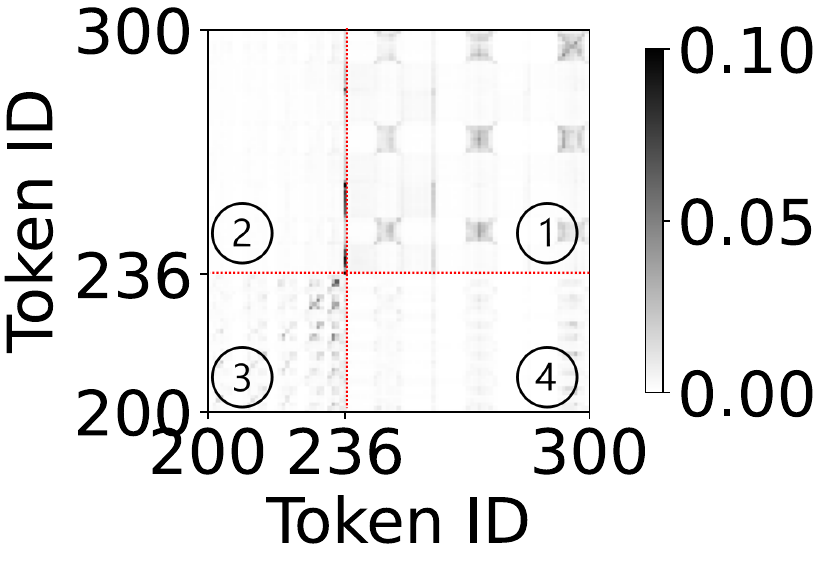}
  \caption{Token activations and attention scores in a SDXL model. \textbf{Left}: Cosine similarity of activations. \textbf{Right}: Zoomed-in visualizaiton of attention scores. Tokens with ID 200-236 are masked and with ID 237-300 are unmasked.}
  \label{fig:token_similarity_attn_heapmap}
  \vspace{-.2in}
\end{figure}

However, we observe that the activations corresponding to unmasked tokens in $\mathbf{Y}$ exhibit high similarity across different requests (those lighter gray cuboids in $\mathbf{Y}$ in \figref{fig:transformer}-Top).
This can be interpretable as the unmasked tokens are supposed to be untouched during image editing.
To verify this, we collect the activations from matrix $\mathbf{Y}$ and calculate the average cosine similarities between corresponding masked and unmasked tokens, as shown in \figref{fig:token_similarity_attn_heapmap}-Left.
The results confirm that the activations for unmasked tokens in $\mathbf{Y}$ are indeed highly similar across different images. 
Additionally, in \figref{fig:token_similarity_attn_heapmap}-Right, we further visualize the attention score matrix ($\mathbf{A}$ in \figref{fig:transformer}) and observe that masked tokens primarily attend to other masked tokens (\circledtext{3}), while unmasked tokens predominantly attend to other unmasked tokens (\circledtext{1}). 
Masked and unmasked tokens attend to each other significantly less (\circledtext{2} and \circledtext{4}), which aligns with the findings in~\cite{ootd}.

Driven by the observation, we propose to reduce the computational load associated with unmasked tokens by reusing cached activations, as illustrated in \figref{fig:transformer}-Bottom.
Given $\mathbf{X}$ with some tokens masked, we extract the matrix of masked tokens, project it into $\mathbf{Q}$, and compute an $\mathbf{Y}$ matrix exclusively for the masked tokens, while replenishing cached activations for the unmasked tokens.
Although reusing cached activations may slightly alter the image generation process, analysis in \figref{fig:token_similarity_attn_heapmap} supports the feasibility of the approach. 
Further, our evaluation in \S\ref{sec:eval_end2end_performance} shows that the images generated using cached activations are visually indistinguishable from those produced through the original computation.




The method of selecting masked tokens is analogous to the decoding process in large language model inference, where the prediction of the next token utilizes the $\mathbf{Q}$ matrix of only the newly generated token along with the $\mathbf{K}$ and $\mathbf{V}$ matrices of all present tokens.

\begin{figure}[t]
  \centering
  \includegraphics[width=0.99\linewidth]{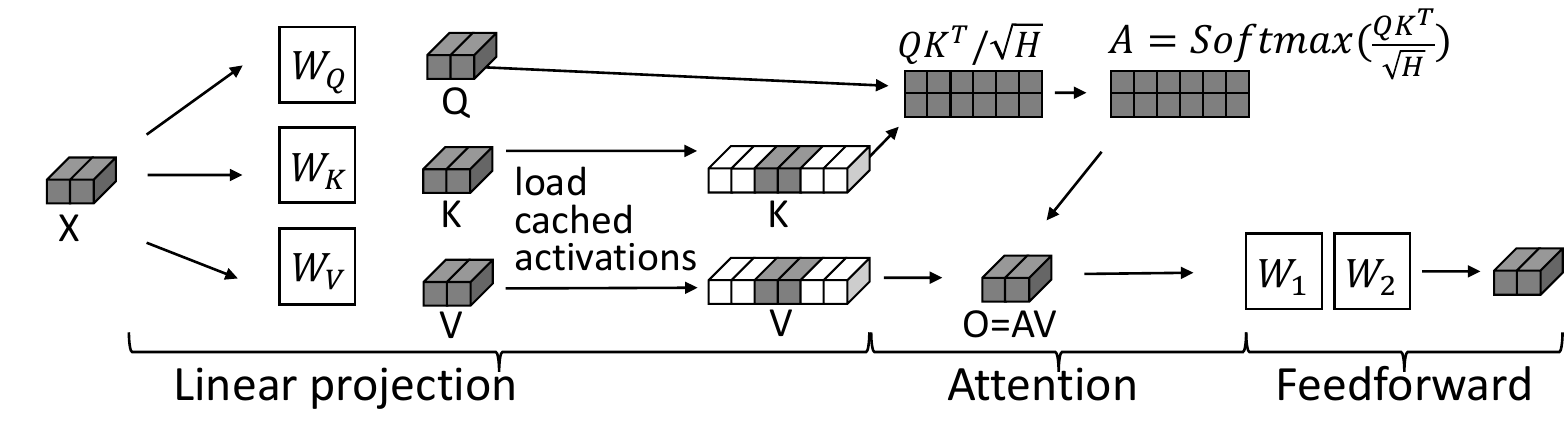}
  \caption{An alternative approach of caching $\mathbf{K}$ and $\mathbf{V}$.}
  \label{fig:transformer_alter}
  \vspace{-.2in}
\end{figure}

\PHB{Alternative approaches.}
While our approach in \figref{fig:transformer}-Bottom utilizes cached activations on $\mathbf{Y}$, an alternative strategy is to apply cached activations on $\mathbf{K}$ and $\mathbf{V}$ instead, as illustrated in \figref{fig:transformer_alter}.
However, this approach doubles the sizes of the cached activations while offering only marginal advantages compared to the approach in \figref{fig:transformer}-Bottom.
With a mask ratio of 20\%, caching $\mathbf{K}$ and $\mathbf{V}$ reduces the latency of a Flux model by 10\% compared to caching $\mathbf{Y}$, from 2.27s to 2.06s.

\subsection{Analysis of Speedup and Caching Overhead}
In this part, we mathematically analyze the speedup and caching overhead associated with the approach in \figref{fig:transformer}-Bottom, as summarized in \tabref{tab:analysis}.
We focus primarily on the computations involved in the feedforward and the linear projection and attention score computations of the attention mechanism.
As indicated in \tabref{tab:analysis},
the computational and caching overhead of mask-aware image editing is predominantly
determined by the batch size $B$ and the mask ratio $m$, since the values of $L$ and $H$ typically remain constant for a given diffusion model.

The serving batch size $B$ within a worker is determined by how a scheduler routes requests across workers and how a worker handles requests in a batch, while the mask ratio $m$ is input-dependent and varies significantly (\S\ref{sec:characterization}).
Consequently, challenges of batch serving and request routing emerge in an image editing serving system (\textbf{C2} and \textbf{C3}).

\begin{table}[t]
    \footnotesize
    \centering
    \def\arraystretch{1.2}
    \begin{tabular}{c | c | c | c |c}
        \hline
        \textbf{}  & {\textbf{Ori. FLOP}} & {\textbf{Acc. FLOP}} & {\textbf{Speedup}} & {\textbf{Cache Shape}} \\
        \hline
        $(\mathbf{XW_1}) \mathbf{W_2}$ & $O(BLH^2)$ & $O(BmLH^2)$ & $\frac{1}{m}$ & $(B, (1-m)\times L, H)$ \\
        \hline
        $\mathbf{XW}$ & $O(BLH^2)$ & $O(BmLH^2)$ & $\frac{1}{m}$ & $(B, (1-m)\times L, H)$ \\
        \hline
        $\mathbf{QK^T} / \sqrt{H}$ & $O(BL^2H)$ & $O(BmL^2H)$ & $\frac{1}{m}$ & $(B, (1-m)\times L, H)$ \\
        \hline
        
        \hline
    \end{tabular}
    \caption{Analysis of the speedup and cache sizes. Without loss of generality, we define an input $\mathbf{X} \in \mathbb{R}^{B \times L \times H}$; a mask ratio $m \leq 1$; 
    two layers in feedfoward computation $\mathbf{W_1} \in \mathbb{R}^{H \times 4H}$ and $\mathbf{W_2} \in \mathbb{R}^{4H \times H}$; $(\mathbf{XW_1}) \mathbf{W_2}$ denotes feed-forward;  $\mathbf{XW}$ denotes linear projection; $\mathbf{QK^T} / \sqrt{H}$ denotes the scaled dot-product attention.}
    \label{tab:analysis}
    \vspace{-.15in}
\end{table}

\section{\sys{} System Design}


\begin{figure}[t]
  \centering
  \includegraphics[width=0.99\linewidth]{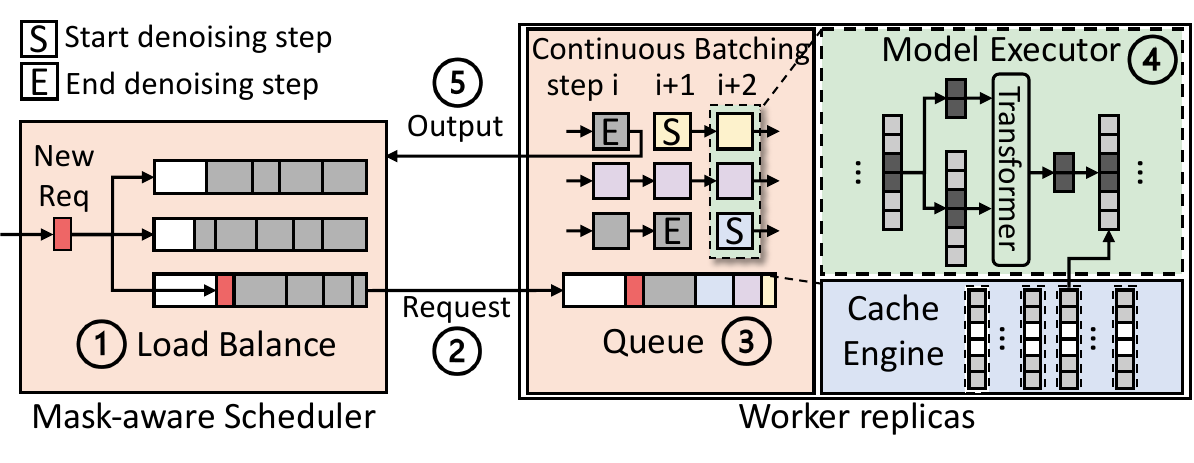}
  \caption{An overall architecture of \sys{}.}
  \label{fig:architecture}
  \vspace{-.2in}
\end{figure}

This section presents \sys{},
an efficient serving system with the \emph{mask-aware} image editing approach (\S\ref{sec:key_insight}).  
Within a worker, \sys{} accelerates the inference of image editing by reusing the pre-computed activations to avoid the redundant computation for the unmasked region. 
Besides, it firstly adapts continuous batching to diffusion model serving to minimize the queueing times of requests.
At the cluster scale, \sys{} features a mask-aware request routing strategy for balancing the workload across workers.

\subsection{System Overview}
\figref{fig:architecture} illustrates the system architecture, which consists of a cluster of worker replicas and a scheduler.

\PHB{Workflow.}
As shown in \figref{fig:architecture}, new requests first arrive at the scheduler (\circledtext{1}), which uses the mask-aware scheduling algorithm described in \S\ref{sec:scheduling} to route them to the appropriate workers (\circledtext{2}). 
Employing a continuous batching strategy detailed in \S\ref{sec:continuous_batching}, the worker fetches requests from the request queue (\circledtext{3}) for the diffusion model (\circledtext{4}) to process, 
wherein the model interacts with the cache engine to speed up image generation through caching, as explained in \S\ref{sec:cache_edit}. 
Finally, the output images are returned to the users (\circledtext{5}).

\subsection{Efficient Image Editing with Caching}
\label{sec:cache_edit}

Following the design approach in \S\ref{sec:key_insight}, we implement an efficient image editing with caching in the \sys{}'s worker replicas. 
Though reusing cached activations can reduce the computational load, the storage and loading of the cached activations pose significant challenges. 
First, the size of cached activations of a template image is large, reaching up to 2.6 GiB for a SDXL model~\cite{ootd}. 
As the number of image templates can be large, storing all activations in the limited high-bandwidth memory (HBM) of GPUs is impractical.
To address this challenge, \sys{} utilizes host memory and disk storage for cached activations storage. 
However, loading cached activations from slow medium to HBM can result in high loading overhead. 
This overhead becomes more significant when the mask ratio is smaller and the size of cached activations gets larger (\S\ref{sec:characterization}).

\begin{figure}[t]
  \centering
  \includegraphics[width=0.99\linewidth]{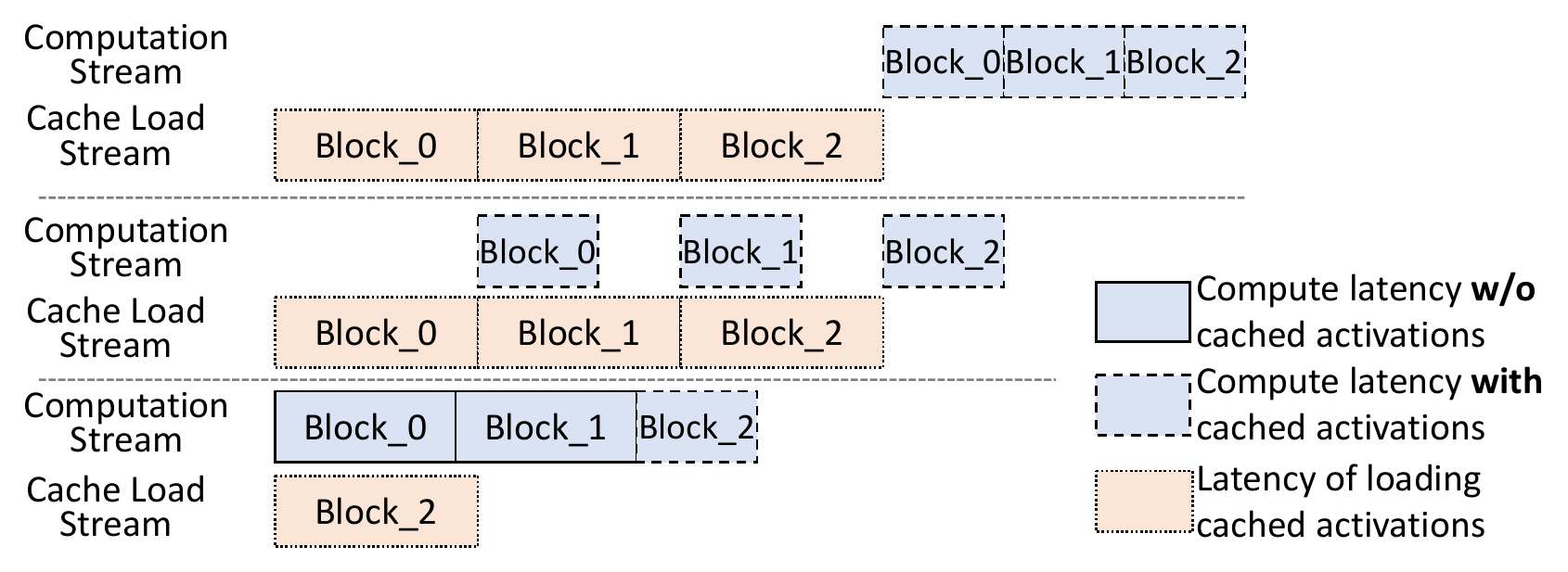}
  \caption{\textbf{Top}: Naive caching loading. \textbf{Middle}: Strawman pipeline loading. \textbf{Bottom}: Bubble-free pipeline loading.}
  \label{fig:pipeline_load}
  \vspace{-.2in}
\end{figure}

\PHB{Strawman solutions to loading cached activations.}
Naively loading cached activations from slower mediums to HBM can block the computation and cause a waste of computational resources, since the computation stream relies on the cached activations to execute the mask-aware image editing inference computation, 
as shown in \figref{fig:pipeline_load}-Top. 
This overhead becomes more significant for the smaller mask ratios~\S\ref{sec:characterization}.

To eliminate this overhead, a strawman solution is to employ a block-wise pipeline loading scheme to mitigate the impact. 
The main idea is to overlap the loading of the cached activations for unmasked tokens with the inference computation of the masked tokens.
In particular, the diffusion model comprises a sequence of transformer blocks, where one transformer's input depends on its precedent's output.
Therefore, we cache the activations at the granularity of transformer blocks. 
As shown in \figref{fig:pipeline_load}-Middle, 
it first loads the cached activations of the first block.
Starting from loading the second block, the  pipeline is built: while loading the $i^{th}$ block, the computation stream can concurrently execute the computation of the $(i-1)^{th}$ block.

However, bubbles will exist in the pipeline.
First, before initiating the computation of the first block, the cached activations for this block must be prepared in the HBM.
The cache load stream first issues the loading of cached activations for the first block into the HBM. 
Only after this is complete can the computation stream start processing the first block.
As a result, a bubble forms due to the loading of the first block by the cache load stream.
Second, when the mask ratio is small, the latency for loading a block can exceed its computation latency. 
In such cases, bubbles will appear between the computations of two adjacent blocks, as shown in \figref{fig:pipeline_load}-Middle.

\PHM{Bubble-free pipeline.}
\sys{} eliminates the pipeline bubbles by selectively using cached activations for transformer blocks within a diffusion model, as illustrated in \figref{fig:pipeline_load}-Bottom.
For blocks that do not use cached activations, \sys{} computes all tokens---both masked and unmasked---without distinguishing between them, thereby avoiding the loading latency associated with cached activations. 
For example, in \figref{fig:pipeline_load}-Bottom, only the activations of $Block_2$ are loaded, while $Block_0$ and $Block_1$ do not use cached activations.
To determine which blocks should use cached activations, we formulate a dynamic programming (DP) problem aimed at minimizing inference latency, as described in Algo.~\ref{alg:dp}.
The complexity of the DP algorithm is $O(N)$, where $N$ is the number of transformer blocks in the diffusion model, typically on the order of tens.
Therefore, the overhead of solving Algo.~\ref{alg:dp} is negligible.



\begin{algorithm}[ht]
\caption{DP for pipeline loading}
\label{alg:dp}
\DontPrintSemicolon
\footnotesize
\KwIn{
$N$: the number of transformer blocks in a diffusion model;
$C_{w.}^m$: the block's computation latency of mask ratio $m$ with cached activations;
$C_{w/o}$: the block's computation latency without any cached activations;
$L^m$: the loading latency of the block of mask ratio $m$.
}
\KwOut{
${useCache}$: a list to indicate whether to use cached activations for each block; ${pipeline\_latency}$: the minimal inference latency of the pipeline.
}
\tcp{\scriptsize Initialize computation \& loading time, caching decisions}
${comp} \leftarrow [0]^{N+1}, {load} \leftarrow [0]^{N+1}, {useCache} \leftarrow [0]^N$ \\
\For{$i \in \{1, 2, \ldots, N\}$}{
    \If{$\max({load}_{i-1} + L^m, {comp}_{i-1}) + C_{w.}^m \leq {comp}_{i-1} + C_{w/o}$}{
        ${load}_i \leftarrow {load}_{i-1} + L^m$\;
        ${comp}_i \leftarrow \max({load}_{i-1} + L^m, {comp}_{i-1}) + C_{w.}^m$\;
        ${useCache}[i - 1] \leftarrow True \quad \vartriangleright \text{Load cached activations}$ \\
    }
    \Else{
        ${load}_{i} \leftarrow {load}_{i-1}$ \\
        ${comp}_{i} \leftarrow {comp}_{i-1} + C_{w/o}$ \\
        ${useCache}_{i-1} \leftarrow False \quad \vartriangleright \text{Compute}$ \\
    }
}
${pipeline\_latency} \leftarrow {comp}_N$ \\
\SetAlgoLined
\end{algorithm}





Algo.~\ref{alg:dp} can also be applied in the case where the mask ratio is large, which means the computation latency for a block with cached activations exceeds the latency of loading those cached activations.
In this case, the inference process becomes computation-bound, and bubbles may appear in the cache load stream.
Despite these bubbles, \sys{} does not eliminate them, as all masked tokens must be processed to ensure image quality.

\PHM{Hierarchical storage for activations}
As we will show in \figref{fig:engine_perf}, the serving throughput of a diffusion model serving engine plateaus at a small batch size of 8, which is typically configured as the engine's maximum batch size~\cite{chen2024Punica}.
Consequently, storing the activations for inflight requests in the running batch usually requires tens of GiBs of host memory, which is negligible compared to the TiB-scale host memory capacities of modern GPU servers~\cite{aws_p4,aws_p5}.
For instance, a machine with 2 TiB of host memory~\cite{aws_p5} can store up to $787$ copies of the activations for the image template used in \figref{fig:edit_example}, providing a sufficiently large cache to accommodate activations of image templates (\S\ref{sec:characterization}).


Despite the capacity of host memory, \sys{} also supports storing cached activations on distributed storage systems or local disks, significantly expanding the storage available for caching activations.
However, the I/O speed of these secondary storage media is on the order of GiB/s, much slower than the tens of GiB/s bandwidth provided by host memory~\cite{katz}.
To utilize the distributed storage system effectively, \sys{} evicts cold activations from host memory to secondary storage based on an LRU (least-recently-used) policy.
When a request arrives, if its required activations are not in host memory, \sys{} begins loading them from secondary storage into host memory. 
This process can run concurrently while the request is queuing, following a state-of-the-practice approach used in KV cache management for LLMs~\cite{cachedattn}.
In \S\ref{sec:eval_end2end_performance}, our evaluation shows that requests often experience a few seconds of queuing time, which is sufficient for loading activations from secondary storage.
For instance, loading the cached activations of the image template in \figref{fig:edit_example} from disk takes 6.4 seconds.



\subsection{Continuous Batching}
\label{sec:continuous_batching}
Leveraging the masks in image editing, \sys{} significantly reduces the computational load per request, which can magnify the performance gain of batching by $1.29\times$ on a Flux model, compared with full-image regeneration.
However, existing diffusion model serving systems often neglect the advantages of batching~\cite{katz, li2024DistriFusion}.
Consequently, these systems typically adopt a simplistic static batching approach~\cite{diffusers_server, anyscale_continuous_batching}, which maintains a fixed running batch size until the running batch completes,
leading to extended queuing times and low GPU utilization.
 


We observe that diffusion models employ an iterative denoising process (\S\ref{sec:primer}), where a latent undergoes multiple denoising steps before being decoded into the final output image. 
The iterative nature of this denoising process is akin to the iterative \emph{decoding} in LLMs. 
Drawing on this parallel, \sys{} adapts the continuous batching strategy to diffusion model serving.
Typically, an image generation request is processed through a sequence of steps: one-step preprocessing, multi-step denoising computations, and one-step postprocessing. 
In diffusion model serving, continuous batching is applied at the step level. 
This means that once a request completes all steps of computation, it is immediately removed from the running batch; new requests can join the batch in just one step, without waiting for the entire batch inference to complete.

\begin{figure}[t]
  \centering
  \includegraphics[width=0.95\linewidth]{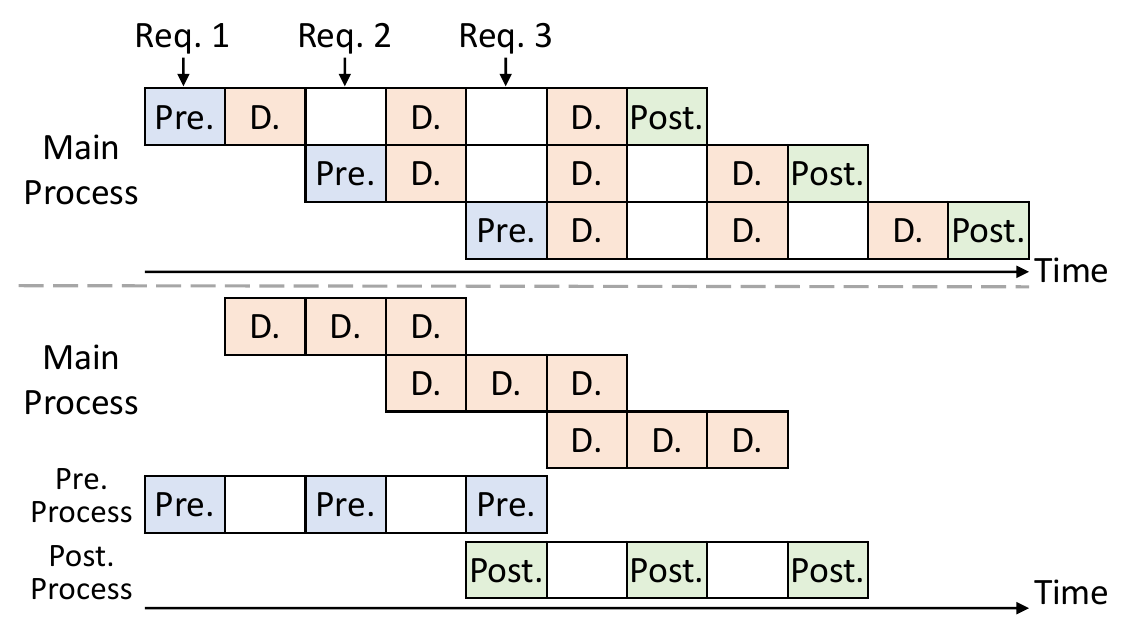}
  \caption{\textbf{Top}: A strawman continuous batching. \textbf{Bottom}: Adapted continuous batching in \sys{}.
\textbf{Pre.}: preprocessing; \textbf{D.}: denoising; \textbf{Post.}: postprocessing.}
  \label{fig:continuous_batching}
  \vspace{-.1in}
\end{figure}

\PHM{Strawman solution.}
A strawman continuous batching is illustrated in \figref{fig:continuous_batching}-Top, where preprocessing and postprocessing can frequently disrupt the denoising computations~\cite{toppings, sarathi}, cumulatively affecting request serving.
In \figref{fig:continuous_batching}-Top, the serving of request 1 is interrupted by the preprocessing of the request 2 and 3.
In diffusion model serving, the preprocessing and postprocessing are CPU-intensive tasks that involve substantial serialization and deserialization computations for images.  
While the overhead of these operations might be negligible when considered individually, their cumulative impact can
significantly increase request serving latency.
As demonstrated in \S\ref{sec:eval_continuous_batch}, our microbenchmark evaluation shows that requests can be interrupted up to 8 times, resulting in a 40\% increase in P95 request serving latency.


\PHM{Disaggregation.}
To address the issue, \sys{} disaggregates the preprocessing/postprocessing from the iterative denoising computation by distributing them across different process, as shown in \figref{fig:continuous_batching}-Bottom.
The main process is dedicated to GPU-intensive denoising computations, while CPU-intensive preprocessing and postprocessing tasks are offloaded to independent processes.
Consequently, the main process will not be interrupted, reducing the tail latency (P95) of request serving by 29\% in evaluation (\S\ref{sec:eval_continuous_batch}).


\subsection{Mask-Aware Scheduler}
\label{sec:scheduling}
Our characterization studies (\S\ref{sec:characterization}) show that masks vary significantly in size.
Therefore, naively scheduling requests across worker replicas, such as using the First-Fit bin-packing algorithm~\cite{chen2024Punica}, will naturally introduce load imbalances for workers.
This issue of load imbalance is also prevalent in LLM serving, where previous research often employs load balancing strategies that assess worker load based on the number of assigned requests or the number of tokens in those requests~\cite{llm_fairness, mooncake}. 
However, these methods fail to accurately gauge the load on a worker in \sys{}, resulting in 35\% increase in request tail latency (P95) in  \S\ref{sec:eval_load_balance}.



\PHM{Estimate a worker's load.}
Based on the design outlined in \S\ref{sec:cache_edit} and \S\ref{sec:continuous_batching}, each worker in \sys{} will handle multiple requests in a batch, which involves \emph{computing} the masked regions for images and \emph{loading} cached activations. 
As shown in \tabref{tab:analysis}, the computational load and cache loading are largely determined by the mask ratio of requests. 
Given the wide variation in mask ratios (\S\ref{sec:characterization}), simple load balancing at the request or token granularity will overlook the impact of mask ratio, failing to accurately reflect the computation and cache loading latencies, which can degrade cluster-level serving performance.


\begin{figure}[t]
  \centering
  \includegraphics[width=0.95\linewidth]{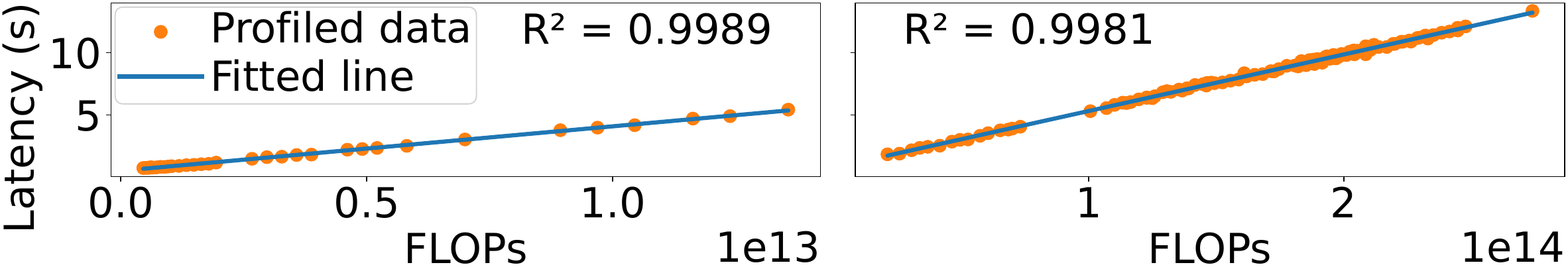}
  \caption{Visualization of the  models to estimate computation latency. \textbf{Left}: SDXL on H800. \textbf{Right}: Flux on H800.}
  \label{fig:regression_model}
  \vspace{-.15in}
\end{figure}

To address the challenge, \sys{} employs linear regression models to estimate computation latency and cache loading latency based on the mask ratios for a batch of requests. 
This approach helps evaluate the load on a worker replica. 
Linear models are chosen because both the computational load and cache sizes scale linearly with the mask ratio (\tabref{tab:analysis}). 
These regression models can be fitted using offline data. 
In \figref{fig:regression_model}, we visualize the models used to estimate the computation latency for a batch of requests.
Each request has different mask ratios.
Following \tabref{tab:analysis}, for a batch of requests, we compute FLOPs of the inference computation based on their mask ratios, which are mapped to inference latency by the regression models.
Our models can accurately fit the data, achieving a high coefficient of determination ($R^2$) of 0.99, \emph{suggesting the models can predict performance almost perfectly}: $R^2$=1 indicates a perfect fit. 
The parameters of the regression models vary with diffusion models and GPUs.


\PHM{Load balance across worker replicas.}
Achieving optimal load balance across worker replicas requires prior knowledge of request details, such as arrival times and mask ratios.
With this information, an optimal load balance schedule could theoretically be available by employing bin packing algorithms to evenly distribute the load among worker replicas. 
However, this assumption is unrealistic in online model serving, where request arrival patterns are bursty~\cite{zhang2023SHEPHERD,gujarati2020Clockwork} and mask ratios vary widely (\S\ref{sec:characterization}). 
Additionally, online migration of requests among worker replicas for load balance~\cite{dLoRA} is impractical in image editing serving due to the significant data communication overhead of large latents, as well as image serialization and deserialization overheads, which can take 20\% of the inference latency of editing an image.

To address these challenges, we utilize the established regression models to develop a \emph{greedy} mask-aware scheduling algorithm that dynamically assigns new requests across worker replicas (Algo.~\ref{alg:scheduling}). 
The scheduler selects the worker replica with the minimum estimated load to handle each new request. 
It keeps track of the runtime status of worker replicas, such as the slack in their running batches. 
Upon receiving a new request, the scheduler identifies candidate workers and calculates a cost score for each one. 
This cost score estimates the load in terms of serving latency on a worker candidate if the new request were allocated to it, derived from extending Algo~\ref{alg:dp},
where the $C_{w_.}^m$, $C_{w/o}$ and $L^m$ of transformer blocks are estimated using the developed regression models. 
The scheduler then assigns the request to the worker candidate with the lowest cost score, ensuring effective load distribution.
In \S\ref{sec:eval_load_balance}, we evaluate our \emph{mask-aware} load balance scheduler, which decreases tail request latency by up to 26\%, compared to baselines.
Additionally, in \S\ref{sec:eval_overhead}, we demonstrate that the load balance scheduler incurs negligible overhead relative to request serving latency.


\begin{algorithm}[ht]
\caption{Mask-Aware Scheduling Policy}
\label{alg:scheduling}
\footnotesize
\SetKwProg{Fn}{Function}{:}{}
\KwIn{$Workers$: a cluster of worker replicas; $R$: a newly coming request; $Comp(\cdot)$, $Load(\cdot)$: lienar regression models for \textit{Computation} and \textit{Cache loading}; \texttt{dp(batch, $Comp(\cdot)$, $Load(\cdot)$)}: a function that extends Algo.~\ref{alg:dp} to return a pipeline and execution latency.}

\Fn{CalcCost{(req, worker)}}{
    \texttt{new\_batch} $\leftarrow$ worker.\texttt{running\_batch} + req \\
    results $\leftarrow$ \texttt{dp(new\_batch, $Comp(\cdot)$, $Load(\cdot)$))} \\
    \Return results.{$pipeline\_latency$}
}

\While{True}{
    \text{Request} $R$ arrives \\ 
    \tcp{\scriptsize Find candidate workers with slack in its running batch}
    candidates $\leftarrow$ available workers \\
    \For{$\text{worker} \in \text{candidates}$} {
        worker.cost $\leftarrow$ \texttt{CalcCost}($R$, \text{worker}) \\
    }
    best $\leftarrow$ $\min$(candidates, key=lambda x: x.cost) \\
    best.serve($R$)
}

\SetAlgoLined
\end{algorithm}

\section{Implementation}
\label{sec:implementation}
\sys{} is an end-to-end serving system featuring a FastAPI frontend~\cite{fastapi}  and a GPU-based inference engine. 
The frontend enables users to customize image generation parameters for requests, including image templates, masks, and input conditions. 
The backend engine is built on Diffusers~\cite{diffusers}, a PyTorch-based diffusion model inference framework that incorporates state-of-the-art model optimization techniques~\cite{diffusers_accelerations}, such as FlashAttn~\cite{dao2023flashattention2}.
Mask-aware image editing is achieved by adapting the attention operator and utilizing CUDA streams to load cached activations asynchronously. 
We implement request queues for continuous batching and load balance scheduler using asyncio~\cite{asyncio}. 
Communication between the scheduler and workers is facilitated via ZeroMQ~\cite{zmq}.
\section{Evaluation}
We evaluate \sys{}'s performance in terms of serving efficiency and image quality. 
We first compare \sys{}'s end-to-end serving performance with strong baselines and then evaluate the effectiveness of \sys{}'s designs, respectively. 
Evaluation highlights include:

\begin{itemize}[topsep=3pt, leftmargin=*, noitemsep, nolistsep, parsep=3pt, partopsep=0pt]

\item \sys{} achieves effcient serving performance while maintaining image quality, reducing request serving latency by $14.7\times$ compared with state-of-the-art baselines (\S\ref{sec:eval_end2end_performance}).

\item \sys{}'s \emph{mask-aware} image editing effectively leverages the sparsity from the mask, achieving empirical results consistent with the theoretical analysis in \tabref{tab:analysis} (\S\ref{sec:eval_mask_aware_edit}).


\item \sys{}'s continuous batching design effectively reduce the queuing times, reducing requests' P95 tail latency by up to  29\%, compared with the static batching and strawman continuous batching solution. (\S\ref{sec:eval_continuous_batch}).

\item \sys{}'s load balance scheduling can descrease the tail request latency by up to 26\% compared to baselines (\S\ref{sec:eval_load_balance}). 

\item \sys{} incurs negligible system overheads (\S\ref{sec:eval_overhead}).  

\end{itemize}

\subsection{Experimental Setup}

\PHB{Models and serving configurations.} 
We use SD2.1~\cite{rombach2022high}, SDXL~\cite{ootd} and Flux~\cite{flux2024} in our evaluation. 
For SD2.1, we serve it with NVIDIA A10 GPUs.
For SDXL and Flux, we serve it on NVIDIA H800 GPUs.
For each model, we use the default settings to generate images, including the denoising steps and image resolutions, for the best image quality.

\PHM{Performance metrics}
Our evaluation mainly concerns two metrics, serving latency and image quality. 
For serving latency, we primarily measure the end-to-end request latency.
For image quality, we use the following quantitative metrics that are widely adopted~\cite{katz, li2024DistriFusion, nirvana, podell2024sdxl, ootd, ootd_dataset,couairon2023diffedit}.

\begin{itemize}[topsep=5pt, leftmargin=*, noitemsep, nolistsep, parsep=3pt, partopsep=0pt]
\item CLIP~\cite{clipscore, clip} score evaluates the alignment between 
generated images and their corresponding text prompts. A higher CLIP score indicates better alignment ($\uparrow$).

\item Fréchet Inception Distance (FID) score~\cite{fid} calculates the difference between two image sets, which correlates with human visual quality perception~\cite{nirvana,katz}. A low FID score means that two image sets are similar ($\downarrow$).

\item Structural Similarity Index Measure (SSIM) score~\cite{ssim} measures the similarity between two images, with a focus on the structural information in images. 
A higher SSIM score suggests a greater similarity between the images ($\uparrow$).

\end{itemize}

\PHM{Baselines.} 
We consider the following baselines.

\begin{itemize}[topsep=5pt, leftmargin=*, noitemsep, nolistsep, parsep=3pt, partopsep=0pt]

\item \Diffusers~\cite{diffusers, diffusers_server} is a standard baseline. 
It uses static batching~\cite{anyscale_continuous_batching, diffusers_server} and does not have a load balance policy. 

\item \FISEdit~\cite{fisedit} accelerates image editing leveraging the sparsity introduced by the mask.
However, it only works with SD2.1 and does not support batching and load balance.

\item \TeaCache~\cite{liu2024timestep} accelerates image generation by caching and reusing intermediate activations to skip computations during the denoising process. 
Although it can be applied to various diffusion models, it suffers from a latency-quality tradeoff.
We configure \TeaCache to minimize its inference latency while ensuring acceptable image quality.
\end{itemize}

\noindent Note that, we implement static batching~\cite{anyscale_continuous_batching} and request-level load balancing for these baselines. 
The advantages of \sys{}'s continuous batching and load balancing will be demonstrated through microbenchmarks in \S\ref{sec:eval_continuous_batch} and \S\ref{sec:eval_load_balance}.

\PHM{Workloads.}
To evaluate online serving efficiency, we generated request traffic following Poisson processes with varying request per second (RPS), which is widely used in  simulating invocations to model serving system~\cite{sheng2023S-LoRA,chen2024Punica, zhang2023SHEPHERD}.
For each request, we set its mask ratio following the distributions in ~\figref{fig:mask_ratio}, which are collected from production traces.
To evaluate the quality of the generated images of each baseline, we include three benchmarks, which are elaborated in~\tabref{tab:image_quality}.



\begin{figure*}[t]
  \centering
  \includegraphics[width=1.0\linewidth]{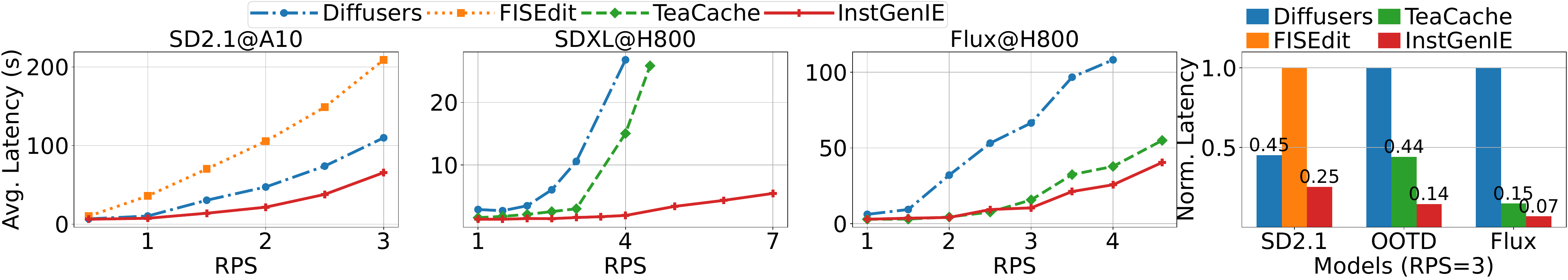}
  \caption{End-to-End request serving performance. \textbf{Rightmost}: Queuing times of requests.}
  \label{fig:end2end_latency}
  \vspace{-.2in}
\end{figure*}

\begin{figure}[t]
  \centering
  \includegraphics[width=0.99\linewidth]{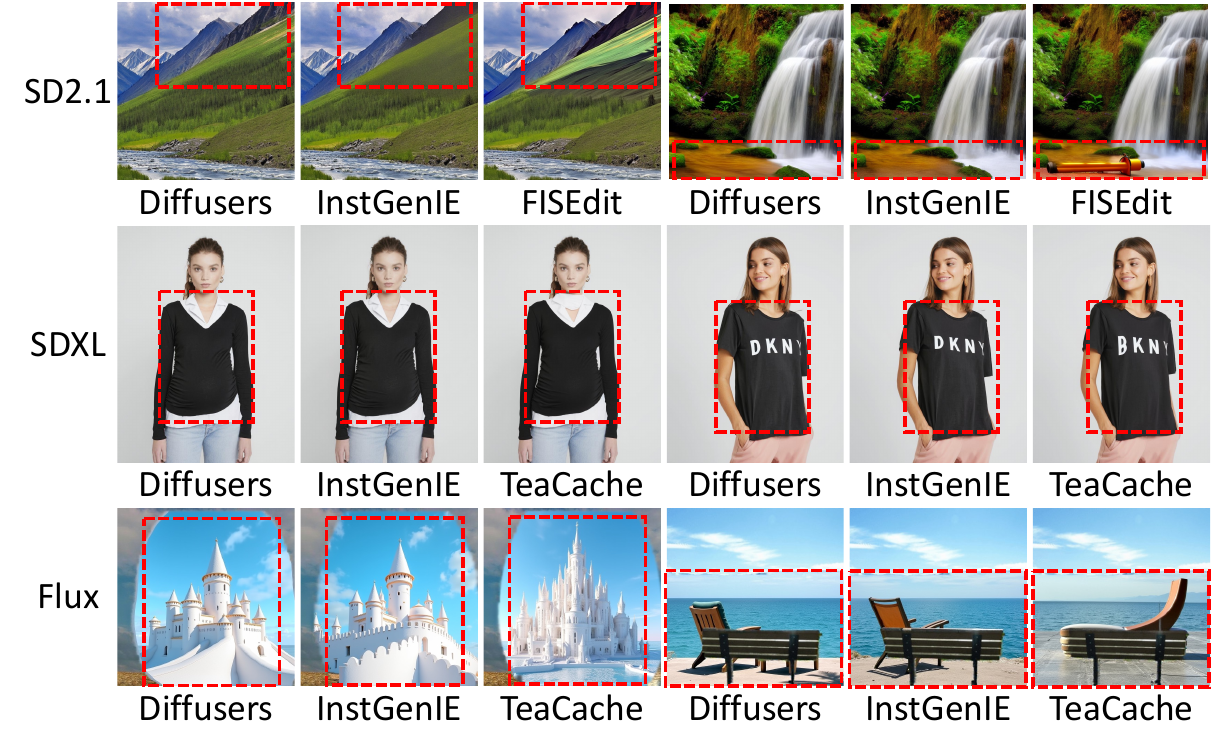}
  \caption{Examples of images generated by each baseline. Major masked areas are circled out with dashed rectangle.}
  \label{fig:sample_images}
  \vspace{-.2in}
\end{figure}

\begin{figure}[t]
  \centering
  \includegraphics[width=0.95\linewidth]{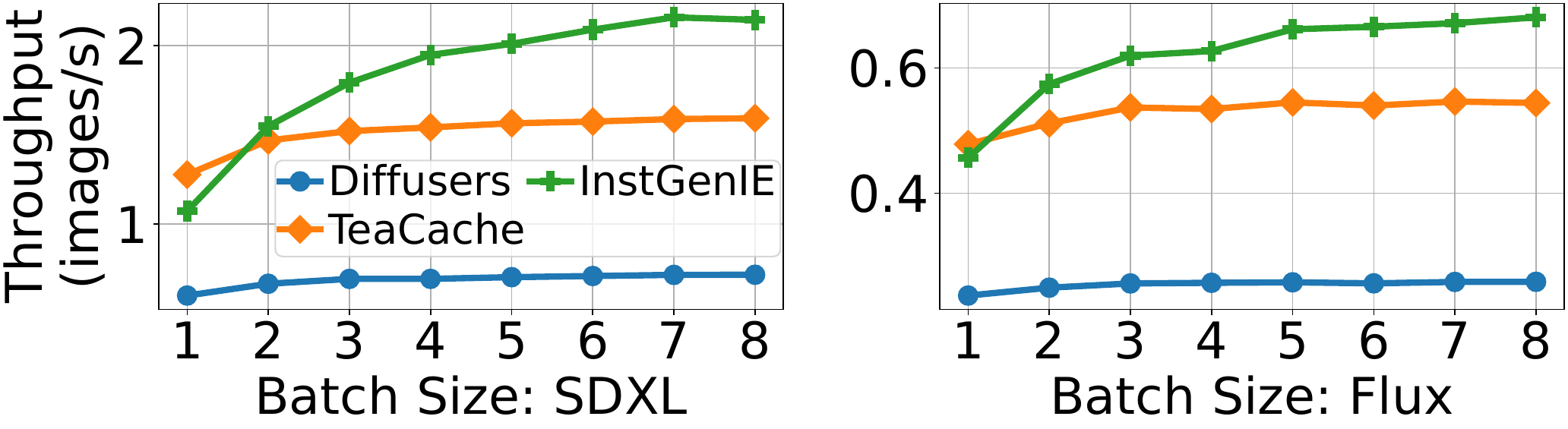}
  \caption{Engine serving performance.}
  \label{fig:engine_perf}
  \vspace{-.2in}
\end{figure}

\subsection{End-to-end performance}
\label{sec:eval_end2end_performance}

\PHB{Online serving efficiency.}
We evaluate the online serving performance on a machine equipped with 8 GPUs, allocating one GPU per worker.
For SD2.1, we use A10 GPUs, while H800 GPUs are used for SDXL and Flux, as \FISEdit is not compatible with NVIDIA Hopper architecture GPUs.
Each baseline is evaluated under varying RPS to shown a spectrum of performance.
The maximum batch size is set to 4 for SD2.1 workers, and 8 for SDXL and Flux.
For each request, we measured its end-to-end serving latency.
As shown in \figref{fig:end2end_latency}, \sys{} consistently outperforms existing systems across all scenarios, reducing the average latency by up to $14.7\times$ compared to \Diffusers, $4\times$ compared to \FISEdit, and $6\times$ compared to \TeaCache.
In the rightmost plot of \figref{fig:end2end_latency}, we present the normalized queuing times for each setting when $RPS=3$.
Compared to the three baselines, \sys{} significantly reduces queuing overhead, thanks to its effective continuous batching strategy (\S\ref{sec:continuous_batching}), leading to more stable serving latencies as RPS increases.
\Diffusers suffers from prolonged model inference latency and substantial queueing overhead because it does not leverage the sparsity introduced by the mask and relies on a static batching policy~\cite{anyscale_continuous_batching} to handle requests.
\FISEdit, on the other hand, does not support batch serving requests with different mask ratios, meaning most requests must be executed one at a time on a worker.
Consequently, requests experience long queuing times, which further exacerate serving latency.
While \TeaCache accelerates model inference, its lack of continuous batching results in considerable queuing overhead.

With the \emph{mask-aware} load balance design (\S\ref{sec:scheduling}), \sys{} also excels regarding tail latency.
At $RPS=3$, \sys{} reduces the P95 request latency by 88\%, 71\%, and 60\% compared to \Diffusers, \FISEdit, and \TeaCache, respectively.

\PHM{Serving engine performance.}
We next evaluate the throughput of each baseline's serving engine under varying batch sizes in \figref{fig:engine_perf}. 
SD2.1 on A10 is omitted because \FISEdit causes GPU OOM errors when the batch size exceeds 2.
Thanks to \emph{mask-aware inference}, \sys{} achieves up to $3\times$ higher throughput than baselines for batch sizes of 2 or larger, featuring a sustained growth in throughput as the batch size increases, whereas the throughput of other baselines plateaus much earlier with marginal batching effects. 

Notably, \sys{} achieves lower throughput than \TeaCache{} without batching (i.e., with a batch size of 1). 
This is due to limited GPU streaming multiprocessor (SM) utilization in \sys{}, as mask-guided selection significantly reduces the number of tokens involved in computation.
In contrast, \TeaCache{} engages all tokens, fully saturating the SMs even without batching.
However, this reduction in token count enhances the effectiveness of batching, necessitating the adoption of continuous batching strategies (\S\ref{sec:continuous_batching}) and helping \sys{} regain its performance advantage in practical serving scenarios where batch sizes are typically large.

\PHB{Image quality.}
We next evaluate the image quality generated by \Diffusers, \FISEdit, and \TeaCache, using \Diffusers as the baseline for generating standard-quality images.
Using three benchmarks, we compare the image quality across these systems and present the results in \tabref{tab:image_quality}.

CLIP scores assess the alignment between generated images and their corresponding textual prompts~\cite{openai_clip}. 
On the benchmarks of SD2.1~\cite{rombach2022high} and Flux~\cite{flux2024}, \sys{} outperforms \FISEdit and \TeaCache, exhibiting better alignment and rivaling \Diffusers’s standard-quality.
For the SDXL benchmark, where input conditions are images (as depicted in \figref{fig:edit_example}), CLIP scores are not applicable.

FID and SSIM scores measure the similarity between the generated images and the standard images (“ground truth”).
Therefore, we use the images generated by \Diffusers as the ground truth, as it represents the standard for diffusion model serving systems.
In \tabref{tab:image_quality}, \sys{} outperforms both \FISEdit and \TeaCache, demonstrating its ability to generate images highly similar to those generated by \Diffusers.
Notably, \sys{} achieves SSIM scores as high as $0.99$, reflecting near-perfect similarity to the images generated by \Diffusers, where the highest possible SSIM score is 1.0.
\figref{fig:sample_images} presents real examples generated by each baseline, where images generated by \Diffusers and \sys{} are visually highly similar, 
while \FISEdit and \TeaCache fail to match the details of \Diffusers.


\begin{table}[t]
    \footnotesize
    \centering
    \def\arraystretch{0.9} 
    \begin{tabular}{@{}c l c c c c@{}}
        \hline
        Model/Dataset        & System  &  CLIP($\uparrow$) &  FID ($\downarrow$) & SSIM ($\uparrow$) \\
        \hline
        \multirow{3}{*}{\thead{ \footnotesize{SD2.1}/ \\ \footnotesize{InstructPix2Pix~\cite{brooks2023instructpix2pix}} } } &  
                        \Diffusers   & 31.4 & - & - \\
                        & \FISEdit   & 31.4 & 50.2 & 0.80 \\
                        & \sys{} (ours)  & 31.8 & 19.9 & 0.92 \\
        \hline
        \multirow{3}*{ \thead{ \footnotesize{SDXL}/ \\ \footnotesize{ VITON-HD~\cite{ootd_dataset}} } }  & 
                        \Diffusers   & - & - & - \\
                        & \TeaCache   & - & 5.4 & 0.97 \\
                        & \sys{} (ours)  & - & 3.4 & 0.99 \\
        \hline
        \multirow{3}*{ \thead{ \footnotesize{Flux}/ \\ \footnotesize{ PIE-Bench~\cite{ju2024pnp}} } }  & 
                        \Diffusers   & 30.9 & - & - \\
                        & \TeaCache   & 30.8 & 77.8 & 0.80 \\
                        & \sys{} (ours)  & 30.9 & 64.8 & 0.88 \\
        \hline
    \end{tabular}
    \caption{Quantitative evaluation on image quality.}
    \vspace{-.15in}
    \label{tab:image_quality}
\end{table}

\begin{figure}[t]
  \centering
  \includegraphics[width=0.49\linewidth]{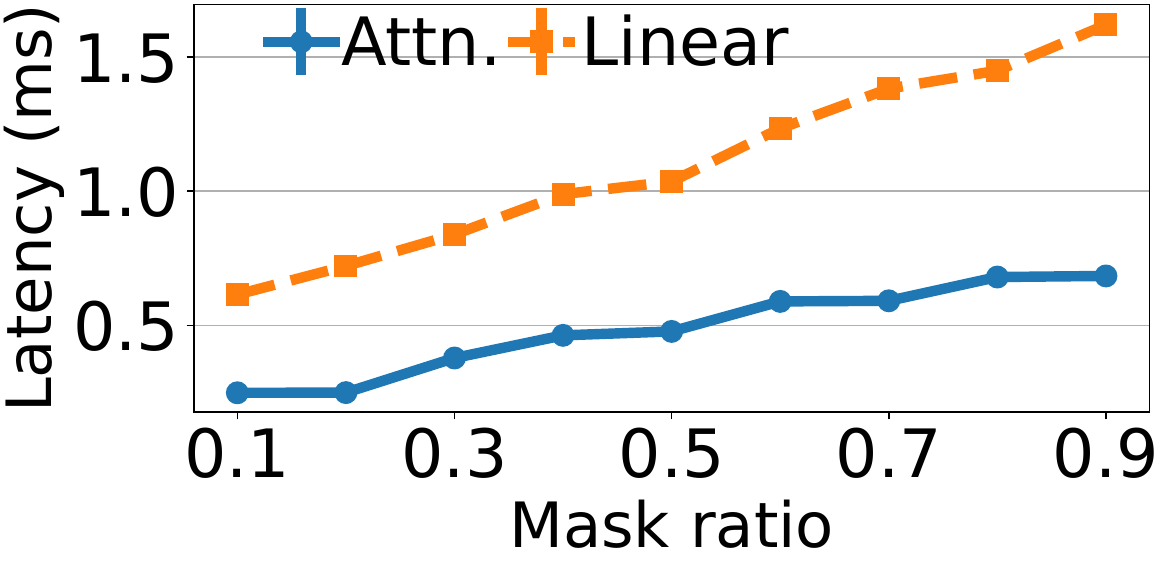}
  \includegraphics[width=0.49\linewidth]{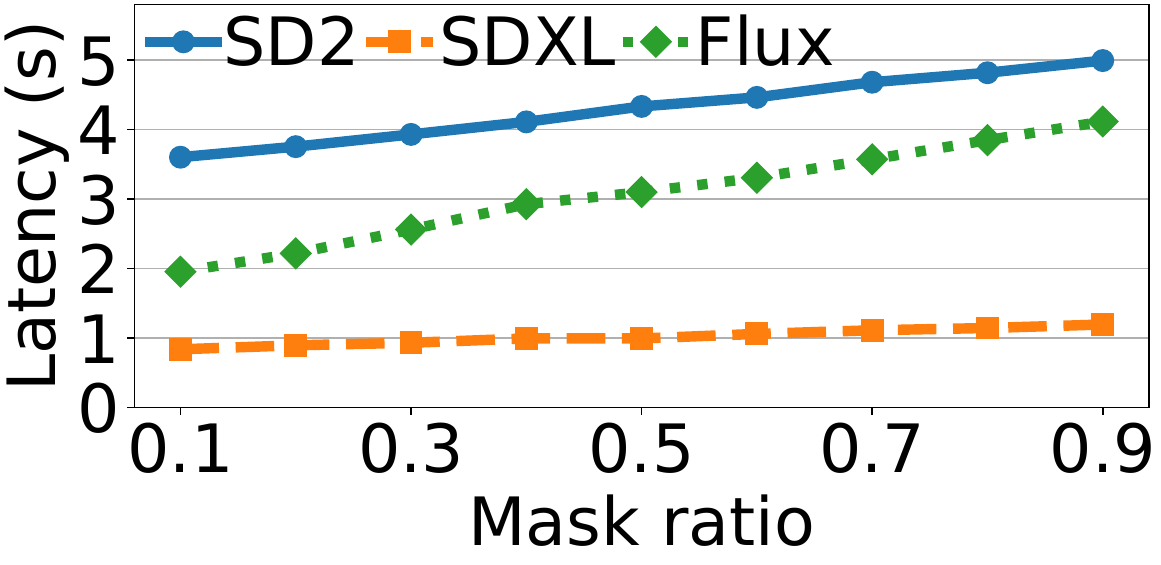}
  \caption{Latency of \emph{mask-aware} image editing with varying mask ratios. \textbf{Left}: Kernel level;
  \textbf{Right}: Image level.}
  \label{fig:micro_kernel_image}
  \vspace{-.2in}
\end{figure}

\subsection{Mask-Aware Image Editing}
\label{sec:eval_mask_aware_edit}

We next evaluate the effectiveness of our \emph{mask-aware} image editing (\S\ref{sec:key_insight}, \S\ref{sec:cache_edit}), which leverages the mask to reduce the computations associated with the unmasked tokens. 

\PHB{Kernel-level performance.} We evaluate the kernel execution latency under varying mask ratios in Flux.
We choose kernels of attention computation and linear computation, the two dominant computations in a transformer block. 
In \figref{fig:micro_kernel_image}-Left, the latency of kernel execution scales linearly with the mask ratio, consistent with the analysis in \tabref{tab:analysis}.

\PHB{Image-level performance.}
We next evaluate the latency of editing an image under different mask ratios using different models.
In \figref{fig:micro_kernel_image}-Right, the latencies of editing an image scale linearly with the mask ratio, consistent with the analysis in \tabref{tab:analysis}.
When the mask ratio is 0.2, \sys{} accelerates the inference with SD2.1/SDXL/FLux by 1.3/2.2/1.9$\times$, by overlaping inference computation with cache loading.


\subsection{Continuous Batching}
\label{sec:eval_continuous_batch}
\begin{figure}[t]
  \centering
  \includegraphics[width=0.49\linewidth]{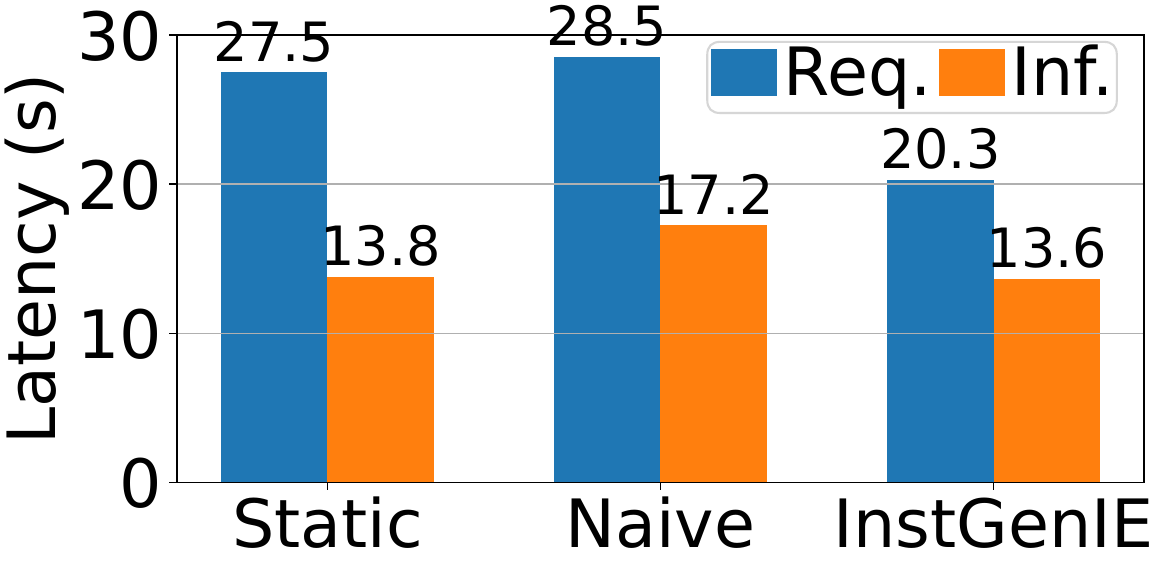}
  \includegraphics[width=0.49\linewidth]{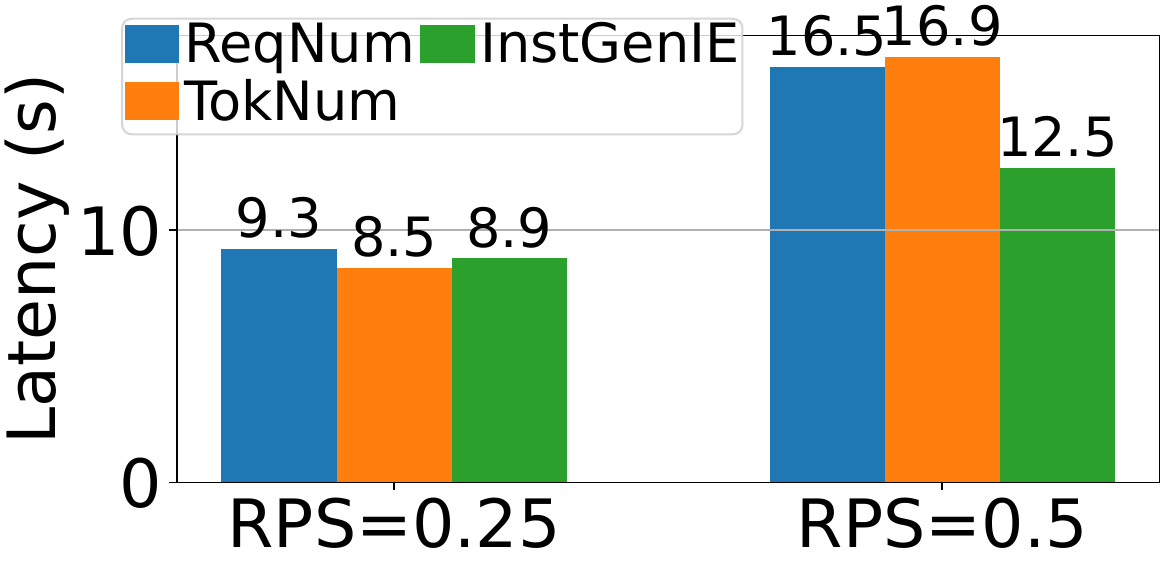}
  \caption{\textbf{Left}: Tail Request (Req.) latency and inference (Inf.) latency using different batching strategies;
  \textbf{Right}: Tail request latency using different load balance policies.}
  \label{fig:micro_cb_lb}
  \vspace{-.2in}
\end{figure}

We next evaluate the benefits of \sys{}'s continuous batching (\S\ref{sec:continuous_batching}). 
We compare the serving performance of a Flux worker with a max batch size of 8 if it adopts static batching~\cite{anyscale_continuous_batching}, naive continuous batching (strawman), and \sys{}'s disaggregated continuous batching, respectively, while other settings are the same.
We measure its performance in terms of P95 tail request latency, using a RPS of 0.5.
\figref{fig:micro_cb_lb}-Left illustrates that staic batching and naive continuous batching can extend the latency by 35\% and 40\%, respectively.
Though requests' inference latency with static batching and \sys{}'s continuous batching are similar, 
static batching degrades because it can incur long queuing latency, where a new arrived requests cannot join in the running batch until the execution of the running batch completes.
Naive continuous batching degrades due to the cumulative interruptions caused by CPU-intensive operations during denoising computations.
The median and P95 interruption times for requests are 6 and 8, respectively. 
Each interruption incurs an average latency overhead of 0.36s, which increases both the inference latency and the overall request latency of each request.

\subsection{Optimizations for Load Balance}
\label{sec:eval_load_balance}
We now evaluate \sys{}'s design for load balance. 
We compare our \emph{mask-aware} method with two baselines: request-granularity load balance and token-granularity load balance.
Essentially, these two methods assign requests solely based on computational load, where request-granularity load balance aims to balance the number of requests assigned to each worker, while token-granularity load balance seeks to balance the number of masked tokens assigned to each worker.
To assess their performance,
we implement each method within \sys{}'s scheduler and 
measure their performance in terms of request tail latency. 
As shown in \figref{fig:micro_cb_lb}-Right, under low request traffic (RPS=0.25) for each worker, the scheduling performances of these methods are comparable, because the overall load on the system is manageable, allowing each method to effectively distribute requests without significant contention or resource saturation.
However, with higher request traffic (RPS=0.5) per worker, 
the performance of the baseline methods degrades, leading to an increase in tail latency by up to 35\%.
This degradation occurs because, at higher traffic levels, the baseline load balancing approaches fail to account for the varying computational and cache-loading demands of requests with different mask ratios. 
Consequently, they may lead to uneven distributions of work among workers, causing some workers overloaded while others remain underutilized.

\subsection{System Overhead}
\label{sec:eval_overhead}
In this part, we analyze the system overhead associated with \sys{} when processing a request, identifying three primary sources.
\textbf{First}, when a request arrives at the scheduler, the scheduler will assess the worker status, make a scheduling decision, and route the request to the appropriate worker, incurring an average overhead of 0.6 ms.
\textbf{Second}, while enabling continuous batching, \sys{} will incur overhead to organize requests' inputs into a batch for denoising computation.
At each denoising step, the batching operation takes 1.2ms on average.
\textbf{Third}, when a worker completes the denoising computations for a request, it should serialize the resulting latent and send it to another process for post-processing. 
The average overhead for serialization is 1.1 ms, while communication adds an additional 1.3 ms.

\PHB{Takeways.} The overhead incurred by \sys{} is on the millisecond scale, which is negligible compared to the overall request processing latency, typically measured in seconds.
\section{Discussion and Related Works}

\PHB{Discussion.} 
\sys{} targets image editing tasks that use masks to specify the editing region. 
As discussed in~\S\ref{sec:characterization}, masks are widely used in  production workloads because they provide precise control over the editing region.
However, for certain image editing tasks, such as style transfer--which modifies the overall appearance of an image--the benefits of mask-aware computation and load balance will diminish.
That said, \sys{}'s continuous batching design is independent of mask usage and can be seamlessly integrated into existing diffusion model serving systems~\cite{diffusers,katz,li2024DistriFusion,nirvana}, enhancing serving performance.

\PHB{Serving diffusion models.}
In \S\ref{sec:existing_works}, we have discussed existing works~\cite{li2024DistriFusion, nirvana, liu2024timestep, diffusers} that are highly related to \sys{}.
Besides, Katz~\cite{katz} focus on accelerating diffusion model inference with adapters~\cite{hu2022lora, zhang2023controlnet}.
Our work is the first to analyze the image editing workloads and addresses the system inefficiencies. 
It is hence orthogonal to the existing optimization solutions for general image generation.

\PHB{Other model serving systems.}
Previous research on model serving systems focuses on optimizing latency~\cite{crankshaw2017Clipper, wang2023Tabi}, throughput~\cite{ahmad2024Proteus},  performance predictability~\cite{gujarati2020Clockwork, zhang2023SHEPHERD}, and  resources efficiency~\cite{zhang2019MArk, wang2021Morphling, gunasekaran2022Cocktail}.
Recent years have witnessed a bloom of LLM serving systems~\cite{wang2023Tabi, yu2022Orca, sarathi, vllm, cachedattn, zheng2024sglang, parrot, chen2024Punica}.
Orca~\cite{yu2022Orca} and vLLM~\cite{vllm} initially introduce continuous batching and apply it in LLM serving. 
A series of works~\cite{cachedattn, zheng2024sglang, parrot} propose to reuse the KV cache of the common prefix prompts to accelerate prefill computing.
Despite their effectiveness in the domain of LLM, directly transfering these techniques to diffusion model serving can lead to suboptimal performance due to the different computation intensity and workflow between LLMs and diffusion models, as discussed in \S\ref{sec:continuous_batching}.


\section{Conclusion}
\label{sec:conclusion}

We presented \sys{}, an efficient system for generative image editing. 
\sys{} effectively leverage the sparsity introduced by masks and 
proposes three novel designs: (1) an efficient pipeline of computing and cache loading to accelerate inference while maintaining image quality (2) a tailored continuous batching for diffusion models, and (3) a mask-aware load balance policy to route requests. 
Collectively, these designs accelerate the inference of image editing and improve the cluster-level serving performance.
Compared to
existing systems, \sys{} achieves $3\times$ higher throughput and reduces average request serving latency by up to $14.7\times$ while maintaining image quality.

\bibliographystyle{ACM-Reference-Format}
\bibliography{main}

\end{document}